\documentclass[12pt,preprint]{aastex}

\usepackage{subfigure,lscape}
\input{psfig.sty}

\shorttitle{ROSAT-HRI investigation of NGC 507}
\shortauthors{Paolillo et al.}

\begin{document}

\title{ROSAT-HRI investigation of the NGC 507 X-ray halo}
\author{M. Paolillo\altaffilmark{1,2}, G. Fabbiano\altaffilmark{3}, G. Peres%
\altaffilmark{1}, D.-W. Kim\altaffilmark{3}}

\altaffiltext{1}{Universit\`a di Palermo - DSFA - Sez.di Astronomia, P.zza del Parlamento 1, 90134 Palermo; paolillo@astropa.unipa.it, peres@astropa.unipa.it}
\altaffiltext{2}{Osservatorio Astronomico di Capodimonte, Via Moiariello 16, 80131, Napoli - ITALY}
\altaffiltext{3}{Harvard-Smithsonian Center for Astrophysics, High Energy Division, 60 Garden St., Cambridge, MA 02138; pepi@head-cfa.harvard.edu, kim@head-cfa.harvard.edu}

\begin{abstract}
We present an X-ray investigation of the elliptical galaxy NGC 507 in the
Pisces cluster. We make use of archival Rosat HRI and \textit{Chandra} data,
and of previously published PSPC data, to connect the large-scale structure
of the halo to the core morphology. Our analysis, based on a bidimensional
double Beta model of the halo surface brightness, shows that the halo core ($%
r<2-3 r_e$) and the external halo ($r>3 r_e$) are characterised by different
dynamical properties and suggests a different origin of the two components.
The halo core has a complex morphology with a main X-ray emission peak,
coincident with the center of the optical galaxy, and several secondary
peaks. The spatial and spectral analysis of the central peak shows that this
feature is produced by denser hot gas in the galaxy core. While both
homogeneous and inhomogeneous cooling flow models predict a deposited mass
exceeding the observed amount, our data support the scenario where the gas
is kinetically heated by stellar mass losses. Comparison with previously
published studies suggest that the core of X-ray extended galaxies is
associated to the stellar distribution and has similar properties to the
X-ray halo of compact galaxies.
The secondary peaks are due instead to
interactions between the radio-emitting plasma and the surrounding ISM,
producing density fluctuations in the hot gas. We found that the energy 
input by the central radio source in the ISM is large enough to prevent gas
cooling and may explain the failure of the cooling flow models.

The total mass profile derived from the bidimensional model shows that a
significant amount of dark matter is present at large radii. The dark halo
extends on cluster scales and is likely associated with the whole cluster
rather than with NGC 507. This structure is typical of many X-ray bright Early-Type
galaxies and may explain the spatial and spectral differences with X-ray
compact galaxies largely debated in literature. The large-scale surface
brightness distribution is irregular and more extended in the North-East
direction. The displacement of the cluster halo from the optical galaxy and
the filamentary structures observed in the halo core further suggest that
the galaxy may be slowly moving within the group potential.

Finally we found that $\sim 20\% $ of the sources detected by \cite{Kim95} 
in the NGC 507 halo are due to point sources while the nature of the remaining 
population is not clear. If associated with NGC 507, they can be either
accreting binaries hosting a massive black hole or density clumps of the
X-ray halo.
\end{abstract}

\section{INTRODUCTION}

NGC 507 is the brightest member of a group of galaxies belonging to a poor
system: the Pisces cluster. The latter in turn is part of the Perseus-Pisces
supercluster \citep[e.g.][]{Wegner93,Sakai94}, located at $\sim 66$ Mpc
\footnote{Assuming H$_0=75$ km s$^{-1}$ Mpc$^{-1}$} from our galaxy \citep{Huchra99}.
NGC 507 is classified as E/S0 galaxy \citep{deVau91}. \cite{Gonzalez00}
showed that, while the surface brightness profile follows a \textit{de
Vaucouleur} law out to $\sim 15"$ from the galaxy center, at larger radii
there is significant excess emission over the $r^{1/4}$ profile. \cite
{Barton98} studied the galaxy distribution in the surrounding of NGC 507
finding that it is part of a group including $\sim 70$ members.

The first studies in the X-ray band based on \textit{Einstein} data %
\citep{Kim92,KFT92} measured a luminosity comparable to those of poor
clusters ($L_X\simeq 10^{43}$ ergs s$^{-1}$) with a temperature $kT>1.5$ keV
at the 90\% level, and suggested the presence of a cooler core. These
findings where confirmed by the Position Sensitive Proportional Counter (PSPC)
on the ROSAT satellite which revealed extended X-ray emission out to
a radius of at least 16 arcmin. \cite{Kim95} were able to resolve the cooler
region within the central 150", thus supporting the presence of a massive
cooling flow, and suggested that cooling clumps may be distributed at large
radii. They estimated a hot gas temperature $\sim 1.1$ keV with no sign of
excess absorption over the galactic value. The data constrained the
metallicity to be not higher than the solar one. \cite{Matsumo97} exploited
the higher spectral resolution and larger energy range of the ASCA satellite
to separate two distinct components in the NGC 507 X-ray spectrum: a soft
thermal emission with $kT\simeq 1$ keV and a harder component with $kT\sim
10 $ keV probably due to unresolved discrete sources. The ASCA data seem to
favour a subsolar metallicity of $\sim 0.3~Z_\odot$. The different metal
abundance obtained by ROSAT and ASCA is explained in part by the new
analysis performed by \cite{Buo00a} on the ROSAT PSPC data, which found a
steep gradient in the metallicity profile of the hot halo.

NGC 507 is a radio galaxy with a steep radio spectrum and weak core %
\citep{Colla75,deRuiter86,parma86}. It may be a source with particularly
weak jets, or possibly a remnant of a radio galaxy whose nuclear engine is
almost inactive and whose luminosity has decreased due to synchrotron or
adiabatic losses \citep{Fanti87}. \cite{Canosa99} used the ROSAT HRI data to
study the correlation between the radio and X-ray emission of a possible
nuclear source, obtaining an upper limit for the nuclear X-ray flux which
exceeds the expectation based on the radio data by more than two orders of
magnitude.

In this work we use data collected by the ROSAT HRI to study the structure
of the NGC 507 hot halo. Our aim is to relate the large scale properties of
the gaseous halo to the small scale structure of the nuclear region. To this
end the HRI is an ideal instrument since it combines a large field of view
with a good angular resolution. We complete our analysis using the recent
Chandra ACIS-S observation to confirm the ROSAT results and to perform a
spectral analysis of the halo core. Our results are compared to those
obtained by other authors on several X-ray bright galaxies in the attempt to
explain the different properties of X-ray extended and compact galaxies.%
\newline
Throughout this work we adopt $H_0=75$ km s$^{-1}$ Mpc$^{-1}$.

\section{OBSERVATIONS AND DATA ANALYSIS}

\subsection{The Data}

The NGC 507 field, including NGC 499, was observed on January 1995 with the
ROSAT HRI \citep{Dav96} for a total time of $\sim 28$ ks (Table \ref
{ngc507observations}). The data, consisting of 2 Observing Intervals (OBI),
were reduced with the SASS7\_5 versions of the ROSAT standard analysis
software (SASS). For our data analysis, we used the IRAF/XRAY and CIAO
packages developed at the Smithsonian Astrophysical Observatory and at the 
\textit{Chandra} X-ray Center (CXC), and other specific software as
mentioned in the text.

To correct for the errors in the aspect time behaviour \citep{Har99}, which
affects the HRI data processed with the SASS versions prior to SASS7\_B of
March 1999 (as it happens for our data), we run the correction routine
ASPTIME (F. Primini 2000, private communication).

\begin{figure}[]
\centerline{\psfig{figure=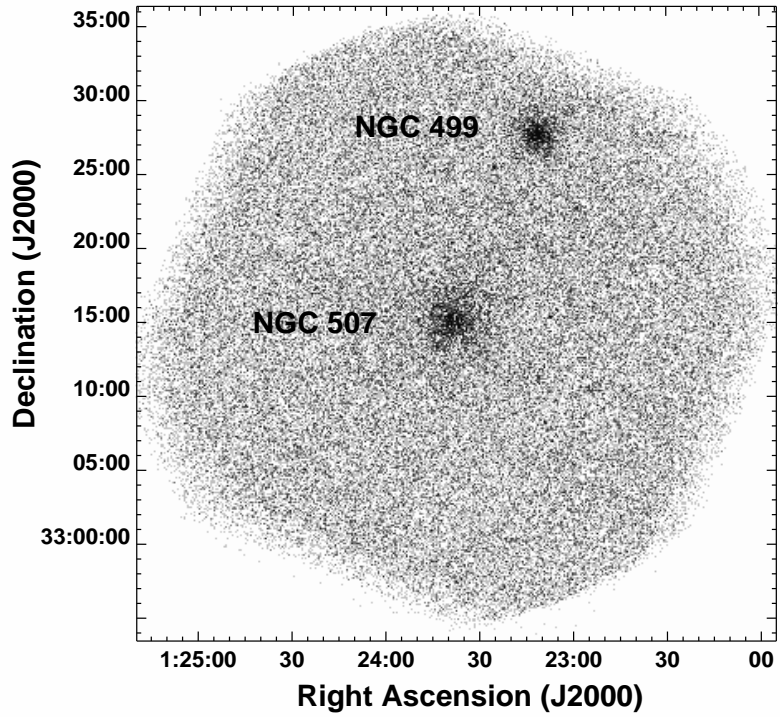,width=0.75\textwidth}}
\caption{Raw HRI image of the NGC 507/NGC 499 field. The data are rebinned
in $5\times 5$ arcsec pixels. North is up and East is to the left.}
\label{ngc507}
\end{figure}

We further tried to correct for problems related to the spacecraft wobble %
\citep{Harris98}; however the pointlike sources detected in our field ($\S$ 
\ref{ngc507_sources}) are not bright enough to allow an accurate
determination of the X-ray centroid (crucial for the correction), when
divided in time bins. Since the global PRF is $\sim 7"$ FWHM we decided to
use the uncorrected data. The HRI field of view (FOV) is shown in Figure \ref
{ngc507}.

We have run the software developed by \citet[ hereafter SMB]{Snow94}, to
produce an exposure map and a background map for our observation. We then
corrected for exposure time and quantum efficiency variations across the
detector by dividing the raw data by this exposure map.

\subsection{Surface Brightness Distribution}
\label{ngc507_brightness} 
\begin{figure}[]
\vspace{1cm} \centerline{\psfig{figure=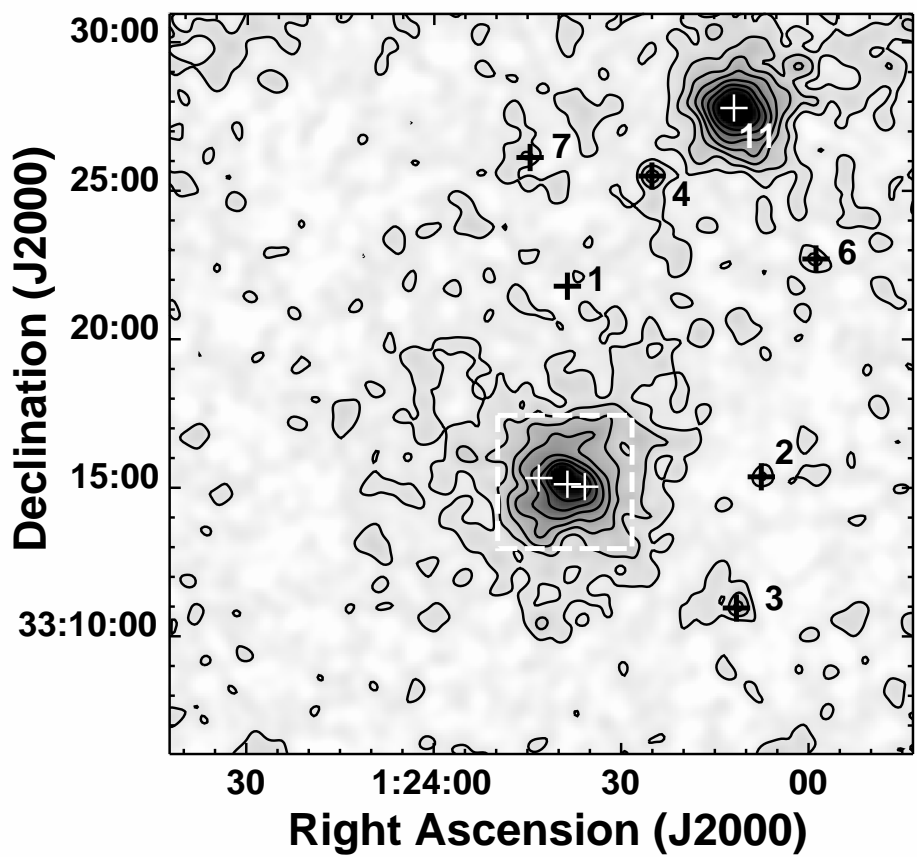,width=0.85\textwidth}}
\caption{Exposure corrected image of the center of the HRI field. The data
are rebinned in $5\times 5$ arcsec pixels and smoothed with a gaussian
filter of $\protect\sigma=15"$. The X-ray contours are spaced by a factor of
1.2 with the lowest one at $6.4\times 10^{-3}$ cnts arcmin$^{-2}$ s$^{-1}$.
Crosses show the position of sources detected by the wavelets algorithm
(Table \ref{ngc507_source_tab}). The white dashed box is the region enlarged
in Figures \ref{ngc507_centbox}, \ref{ngc507_radio} and \ref{ngc507_model}.
In the figure North is up and East is to the left.}
\label{ngc507_smooth}
\end{figure}

To study the surface brightness distribution of
NGC 507 the exposure corrected data were rebinned in $5\times 5$ arcsec
pixels and then convolved with a gaussian filter of $\sigma=15"$ (Figure \ref
{ngc507_smooth}). The resulting image shows the presence of an extended
X-ray halo surrounding the galaxy. The central part of the halo (the one
enclosed in the dashed box) is relatively symmetric but elongated in the
NE-SW direction. Two filamentary structures seem to depart from the galactic
center toward NW and SE respectively. The morphology of the outer halo ($%
r>200"$) is very irregular due to the low S/N ratio of the HRI image.
However there appears to be more emission on the North and East sides with
respect to the South and West ones. This is confirmed by the radial profiles
discussed in $\S$ \ref{ngc507_brightness_prof}. In contrast with NGC 507,
the NGC 499 halo is less extended and much more symmetric. The positions of
the sources detected by the wavelets algorithm ($\S$ \ref{ngc507_sources})
and discussed in $\S$ \ref{ngc507_sources}, are marked with crosses.

\begin{figure*}[p]


\centerline{\psfig{figure=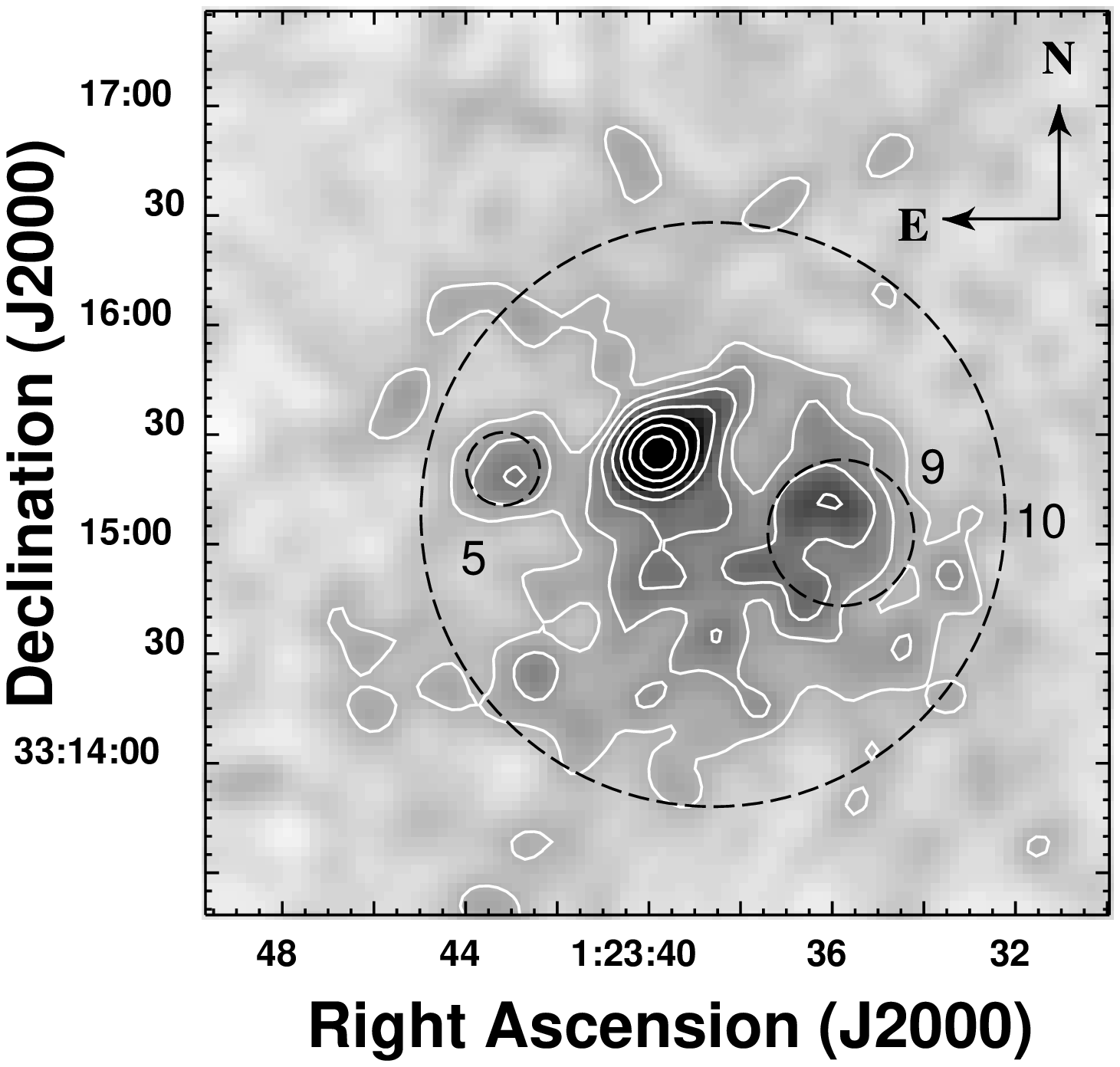,angle=0,width=0.53\textwidth}
\psfig{figure=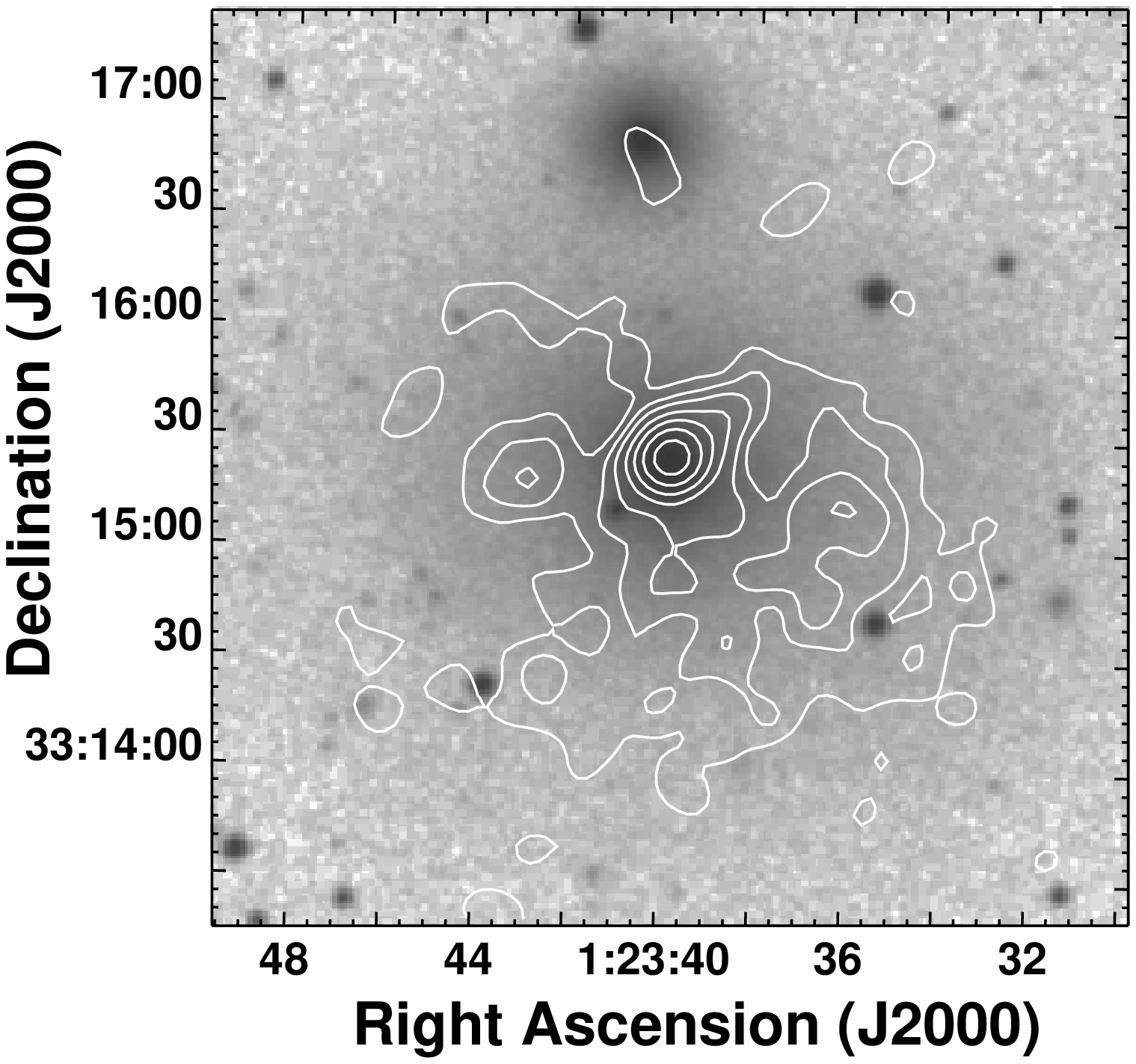,angle=0,width=0.52\textwidth}}
\caption{\textbf{\textit{Left:}} $2.5\times 2.5$ arcsec/pixel image of the
nuclear region of NGC 507. The image is convolved with a gaussian filter of $%
\protect\sigma=5"$. Contours are spaced by a factor 1.3 with the lowest one at 
$1.4\times 10^{-2}$ cnts arcmin$^{-2}$ s$^{-1}$. The dashed circles show the 
region of maximum S/N ratio for the sources detected by the wavelets algorithm. 
A higher quality color version of this figure is available in the electronic 
edition. \textbf{\textit{Right:}} X-ray brightness contours overlaid on the 1 
arcsec/pixel Digital Sky Survey (DSS) image (logarithmic grayscale).}

\label{ngc507_centbox}
\end{figure*}

The detection of three distinct emission peaks (white crosses) in the core
of the X-ray halo (dashed box in Figure \ref{ngc507_smooth}) suggests a
complex structure of the nuclear region. Examining this area in more detail
(left panel of Figure \ref{ngc507_centbox}), a number of interesting
features can be detected: 1) a strong emission peak can be observed in the
center of the X-ray halo; this peak is slightly elongated due to the
presence of two `tails' extending in the NW and the SE direction; 2) a
secondary peak is visible $\sim 50"$ West of the primary peak (source No.9);
3) a region of low X-ray brightness is interposed between the primary and
secondary peak; 4) a third peak is visible $\sim 40"$ East of the galaxy
center (source No.5); 5) additional diffuse emission is present on the
Southern side of the nucleus, making the surface brightness gradient flatter
than on the Northern side.

The main peak is almost coincident with the center of the optical galaxy
(Figure \ref{ngc507_centbox}, right panel). The second and third peak,
instead, are not related to any optical feature, although being within the
stellar body of the galaxy. There is also some faint emission coincident
with the position of the elliptical galaxy NGC 508, located 1.5 arcmin North
of NGC 507: this detection has a low statistical significance 
($\sim 2\sigma$) but it is confirmed by the \textit{Chandra} data analysis 
performed in $\S$ \ref{ngc507_chandra}.

\begin{figure}[]


\centerline{\psfig{figure=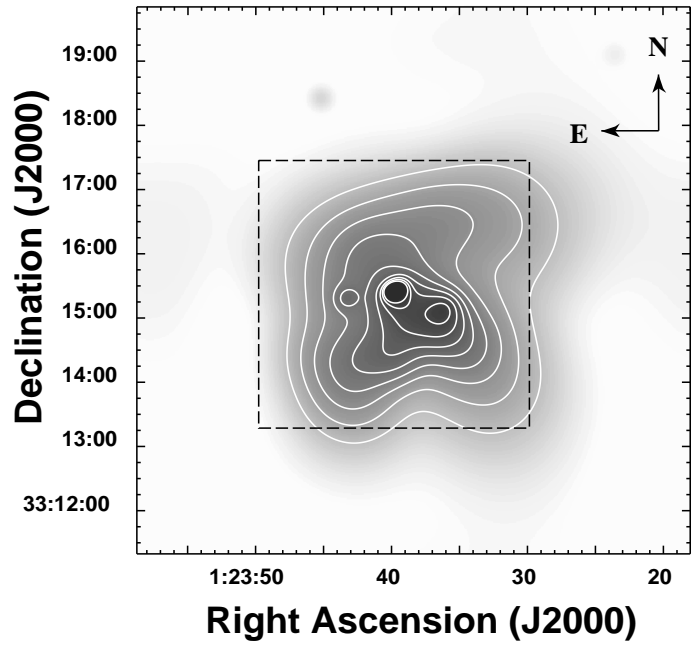,width=0.8\textwidth}}
\caption{Adaptively smoothed image of the center of the NGC 507 halo. The
data are rebinned in $1\times 1$ arcsec pixels. Contours are spaced by a factor 
1.2 with the lowest one at $8.3\times 10^{-3}$ cnts arcmin$^{-2}$ s$^{-1}$. 
The dashed box is the region enlarged in Figures \ref{ngc507_centbox}, 
\ref{ngc507_radio} and \ref{ngc507_model}. A higher quality color version of this figure is available in the electronic 
edition.}

\label{ngc507_csmooth}
\end{figure}

To reveal the presence of extended low S/N features we adaptively smoothed
the HRI image with the \textit{csmooth} algorithm, contained in the CIAO CXC
package, which convolves the data with a gaussian of variable width
(depending on the local signal to noise range of the image) so to enhance
both small and large scale structures. In the resulting image (Figure \ref
{ngc507_csmooth}), in addition to the three emission peaks mentioned before,
we can distinguish an elongated tail extending on the North and bending
Westwards out to $\sim 200"$ (63 kpc \footnote{%
At the the assumed distance of NGC 507, 1'=19.1 kpc}), and an X-ray tongue
protruding by $\sim 160"$ (51 kpc) on the South-East side. These two
features correspond to the filamentary structures visible also within the
dashed box in Figure \ref{ngc507_smooth}.

\subsection{X-ray/Radio Comparison}

\label{ngc507_Xradio} 
\begin{figure*}[]


\centerline{\psfig{figure=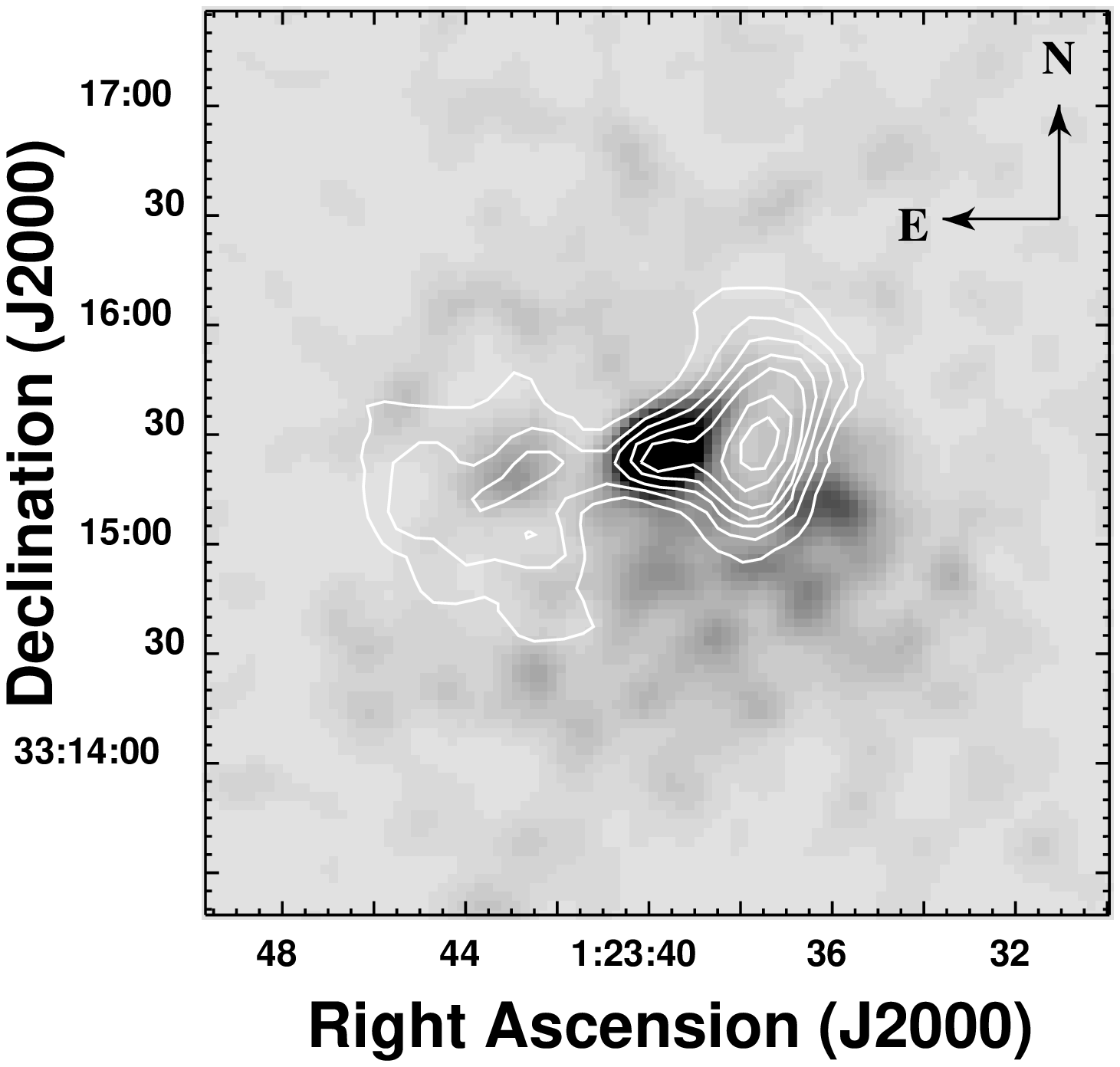,angle=0,width=0.5\textwidth}
\psfig{figure=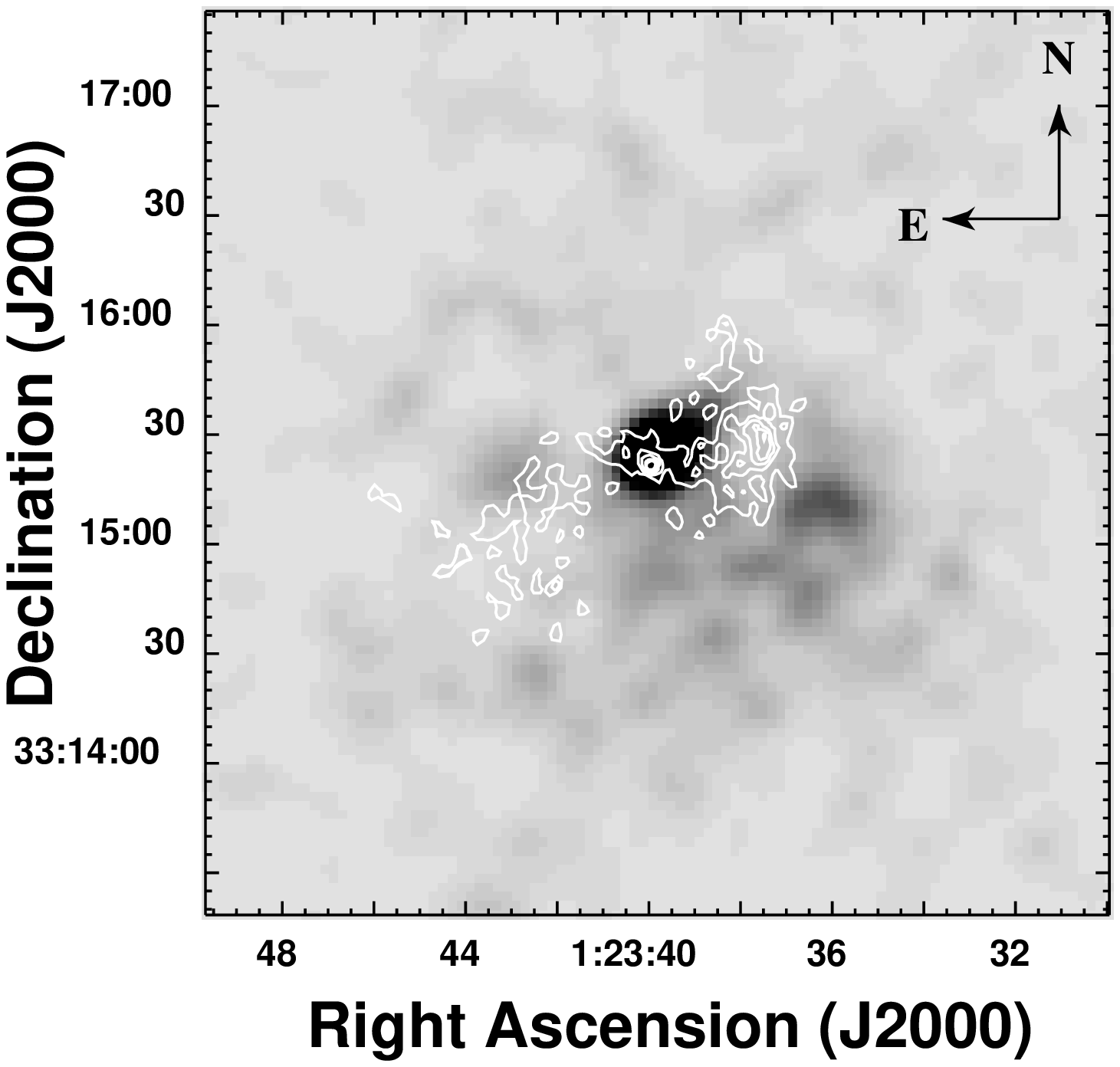,angle=0,width=0.5\textwidth}}
\caption{\textbf{\textit{Left:}} 13" FWHM radio contours superimposed on the
X-ray nuclear region of NGC 507. The X-ray image is the same shown in the
left panel of Figure \ref{ngc507_centbox}. \textbf{\textit{Right:}} Higher
resolution (3.5" FWHM) radio contours superimposed on the X-ray nuclear
region of NGC 507. Higher quality color versions of these figuree are available 
in the electronic edition.}

\label{ngc507_radio}
\end{figure*}

NGC 507 is known to be a weak radio source. In Figure 
\ref{ngc507_radio} we superimpose the radio contours at 20 cm %
\citep{parma86,deRuiter86} on the smoothed X-ray image.

The nuclear radio source, visible in the high resolution map (right panel of
Figure \ref{ngc507_radio}) is coincident with the position of the central
X-ray peak. However it seems to be slightly displaced ($\sim 4"$) toward the
South-East edge of the peak. The radio jets/lobes appear aligned with the
direction of the secondary X-ray peaks. The western jet appears to be
collimated up to 10-15 arcsec ($\sim 4$ kpc) and then expands in a large
lobe. The latter is coincident with the region of low surface brightness
described in $\S$ \ref{ngc507_brightness}. The higher resolution radio
contours (Figure \ref{ngc507_radio}, right panel) reveal a very good
agreement between the morphology of the Western lobe and the shape of the
`cavity'. The secondary X-ray peak lies at the end of the radio lobe.

The correlation between the X-ray and radio surface brightness for the Eastern lobe, which is fainter than the Western one (27 mJy compared to 40 mJy at 1.4 GHz; \citealp{deRuiter86}) is only visible when we look at the high resolution
map: the radio emission seems displaced South of the X-ray peak and falls in
a region of low X-ray emission.

\subsection{Bidimensional Halo Model}

\label{ngc507_2Dmodel} To study the complex structure of the inner halo of
NGC 507 we built a bidimensional model using the \textit{Sherpa} fitting
software contained in the CIAO CXC package. The model was composed by adding
together two bidimensional Beta components of the form: 
\begin{equation}  \label{2dmodel}
\Sigma(x,y)=\frac{A}{(1+(r/r_0)^2)^{3\beta-0.5}}
\end{equation}
where 
\begin{equation}
r(x,y)=\frac{x^2_{new}(1-\epsilon)^2+y^2_{new}}{1-\epsilon}
\end{equation}
and 
\begin{equation}
x_{new}=(x-x_0)\cos(\theta)+(y-y_0)\sin(\theta)
\end{equation}
\begin{equation}
y_{new}=(y-y_0)\cos(\theta)-(x-x_0)\sin(\theta).
\end{equation}
where $x_0,y_0$ are the coordinates of the X-ray centroid, $\beta$ is the
slope of the surface brightness profile, $r_0$ is the core radius, $\epsilon$
is the ellipticity and $\theta$ is the position angle (P.A.). The two
components represent respectively the central X-ray peak and the extended X-ray halo. The 2D model was convolved with the HRI PRF \citep{Dav96} and
then fitted on the exposure corrected HRI image. To model simultaneously small and large scale components, we fitted the extended component on a 
$30\times 30$ arcsec pixel image and the central component on a higher
resolution $5\times 5$ arcsec pixel image.

\begin{figure*}[]


\vspace{1.5cm} \centerline{\psfig{figure=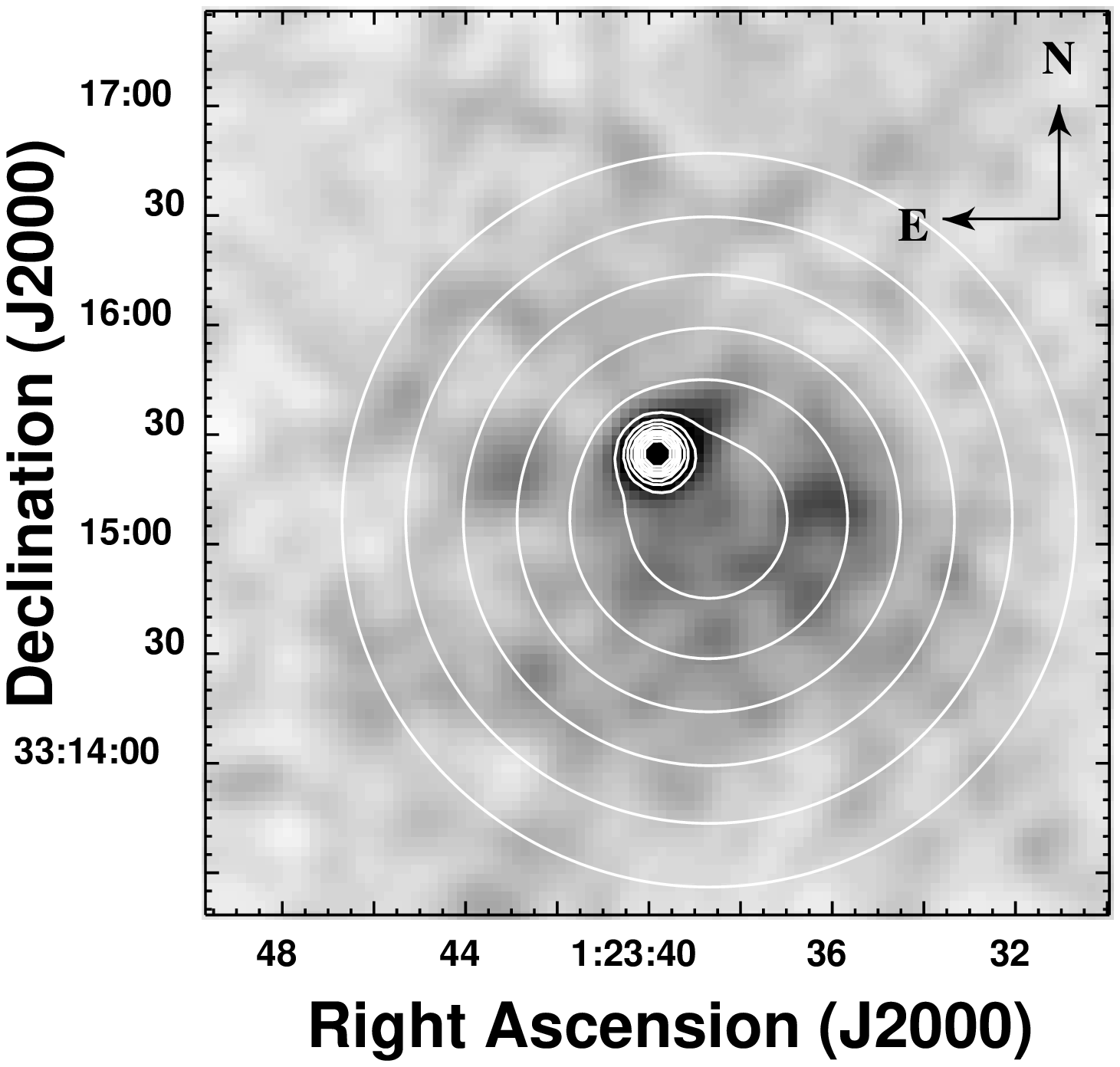,width=0.5\textwidth}%
\psfig{figure=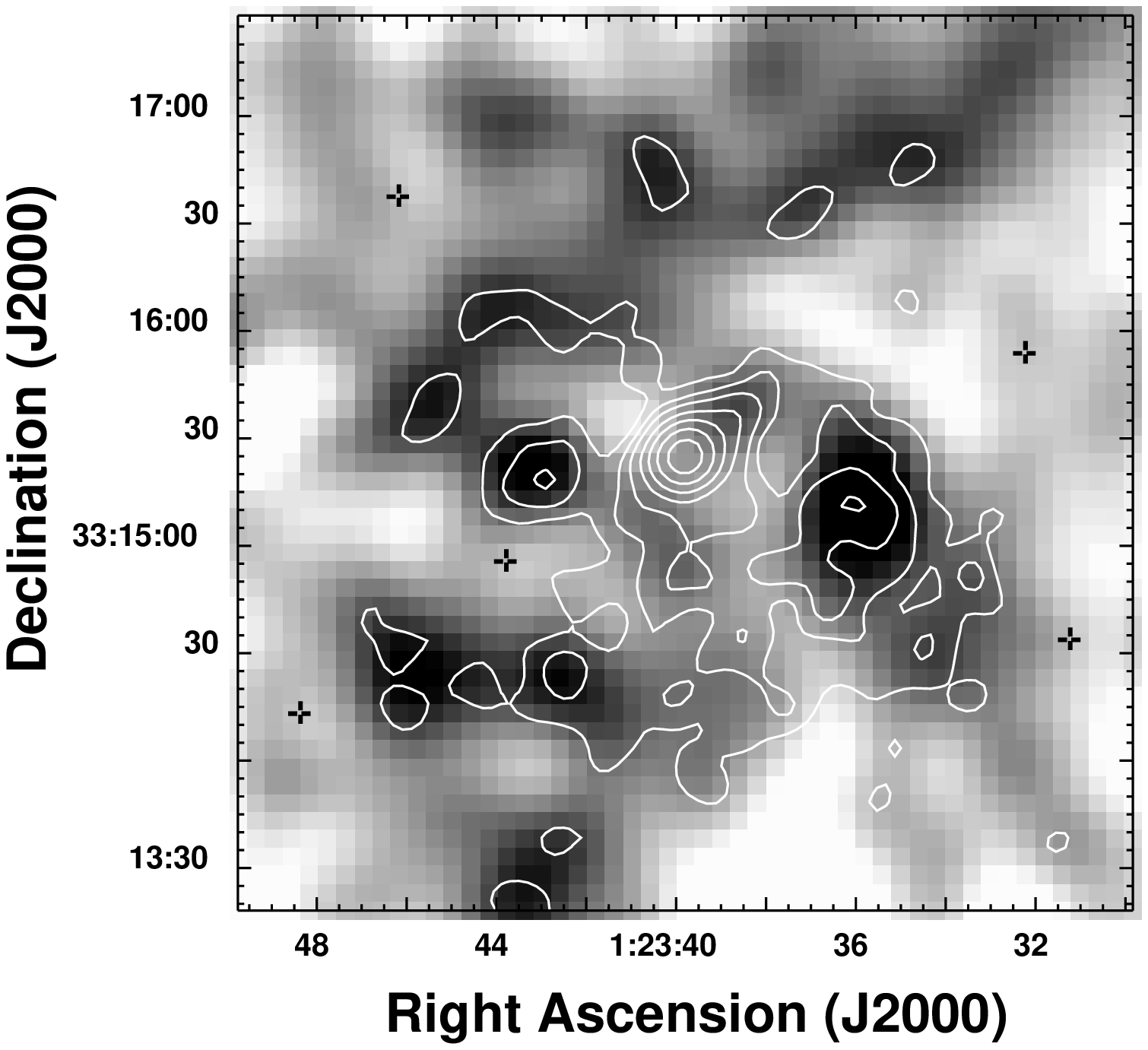,width=0.51\textwidth}}
\caption{\textbf{\textit{Left:}} X-ray contours of the bidimensional model
superimposed on the NGC 507 central halo. The model contours are spaced by a
factor 1.3 with the lowest one at $9.6\times 10^{-3}$ cnts arcmin$^{-2}$ 
s$^{-1}$. \textbf{\textit{Right:}} X-ray contours superimposed on the
residuals of the bidimensional model. The contour levels are the same of the
right panel of Figure \ref{ngc507_centbox}. The residuals range from 
$-1\times 10^{-3}$ (white) to $+2\times 10^{-3}$ cnts arcmin$^{-2}$ s$^{-1}$
(black). The crosses show the grayscale level corresponding to zero. 
Higher quality color versions of these figuree are available 
in the electronic edition.}

\label{ngc507_model}
\end{figure*}

Table \ref{ngc507_fit_tab} shows the best-fit parameters of the
bidimensional model. For the large scale, fit the ellipticity and the P.A.
of the two components were fixed to 0 because of the poor S/N ratio of the 
HRI image at large radii.
For the central component fit leaving these two parameters free to vary
yields $\epsilon=0.56\pm 0.02$, $\theta=137^\circ$, and results in minor
changes on the slope ($\beta=0.63\pm0.01$) and the core radius ($r_0=0.32\pm
0.02$). The central component is thus elongated in the NW-SE direction, as
expected from the presence of the nuclear `tails' ($\S$ \ref
{ngc507_brightness}). The centroid of the extended component is displaced by $\sim
22"$ (7 kpc) South-West of the central emission peak (Figure \ref
{ngc507_model}, left panel); this is in agreement with the position of
source No.10 detected by the wavelets algorithm (left panel of Figure \ref
{ngc507_centbox}) which corresponds to the centroid of the large-scale
X-ray distribution. The best-values $r_0$ and $\beta$ of the extended component are consistent
within 1 $\sigma$ with those found by \cite{Kim95}; however our core radius
is slightly larger due to the fact that we have separated the contribution
of the central peak.

In the right panel of Figure \ref{ngc507_model} we show the residuals of the
bidimensional model. Most of the structure already discussed in $\S$ \ref
{ngc507_brightness} is clearly visible, in particular we can distinguish
the secondary peaks and the X-ray excess, located on the NW side of the
central peak, corresponding to the nuclear `tail'. On the Northern side of
the halo there is an elongated X-ray feature corresponding to the Northern
tail already seen in the adaptively smoothed image of Figure \ref
{ngc507_csmooth}.

\subsection{Radial Brightness Profiles}

\label{ngc507_brightness_prof} 
\begin{figure}[]
\centerline{\psfig{figure=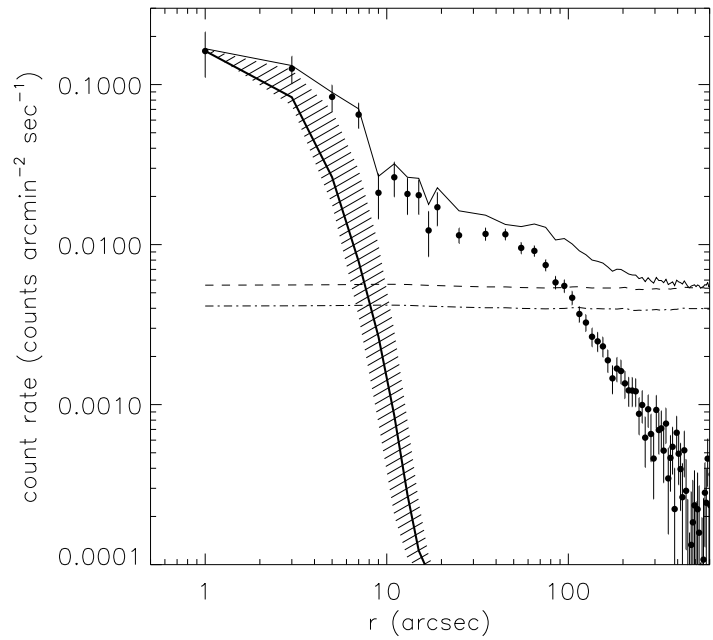,width=0.6\textwidth}}
\caption{HRI surface brightness profile of NGC 507 (thin continuous line).
The dashed (dot-dashed) line represent the SMB background map profile after
(before) rescaling to match PSPC counts. Filled circles show the background
subtracted profile obtained using the rescaled SMB background. The PRF
profile and the uncertainties due to residual aspect errors, are shown
respectively as the thick continuous line and the shaded region.}
\label{ngc507_profile}
\end{figure}

To study quantitatively the hot halo of NGC
507 we derived radial profiles of the X-ray surface brightness. We binned
the X-ray counts in 2 arcsec annuli, centered on the main X-ray peak, up to
20 arcsec and in 10 arcsec annuli at larger radii. This allows to both
exploit the HRI resolution in the inner regions, where the X-ray emission is
stronger, and to have a good S/N ratio in the outer regions. The brightness
profile is shown in Figure \ref{ngc507_profile} as a thin continuous line.

To derive the instrumental background level we used the background map
produced by the SMB software (dot-dashed line in Figure \ref{ngc507_profile}%
). Since the latter may be a poor representation of the actual background
when an extended source is present in the field 
\citep[see for instance
][]{paolillo02}, we compared the HRI data with the PSPC profile obtained by 
\cite{Kim95}. The HRI profile was rebinned in 30 arcsec annuli to match the
PSPC resolution and the SMB background was then rescaled to obtain the best
agreement between the two profiles in the 100-500 arcsec range. The final
adopted background (dashed line) is in close agreement with the flattening
level of the HRI profile (i.e. the brightness level measured at radii $> 500$
arcsec), thus indicating that the measured background is dominated by the
instrumental contribution expected when no source is present. Using this
background determination we derived the background-subtracted profile shown
as filled circles.

The inner 10 arcsec are dominated by the emission due to the central X-ray
peak shown in left panel of Figure \ref{ngc507_centbox}. Even though a
comparison with the HRI PRF suggests that the peak is extended, taking into
account the PRF uncertainties due to residual aspect problems (shaded region
in Figure \ref{ngc507_profile}; \citealp{Dav96}) and the asymmetries
discussed below, we find that the central peak is marginally consistent with
the presence of a pointlike nuclear source. However the analysis of the 
\textit{Chandra} data ($\S$ \ref{ngc507_chandra}) performed by \cite
{Forman01a} unambiguously show that the central component is extended.

Past 20 arcsec ($\sim 6$ kpc) the X-ray profile flattens due to the presence
of the extended component, and then decreases with a power-law profile
for $r>60"$.

Figure \ref{ngc507_model_profile} shows the model profile ($\S $ \ref
{ngc507_2Dmodel}) -- convolved with the HRI PRF -- superimposed on the HRI
data. The contribution of the two components are shown by the dashed and the
dot-dashed line for the central and extended component respectively. The
PSPC profile measured by \cite{Kim95} is in very good agreement with the HRI
data and shows that the halo brightness decreases as a $r^{-1.88}$ power-law
even outside the range covered by the HRI points. For comparison we also
show the optical V band surface brightness profile measured by \cite
{Gonzalez00}. We can see that in two regions, for $10"<r<20"$ and $%
40"<r<100"$, the HRI counts are in slight excess with respect to those
predicted by the bidimensional model. These excesses are due respectively to
the presence of the nuclear tails discussed $\S $ \ref{ngc507_brightness}
and to the X-ray emission at the edge of the SW radio lobe.

We have shown in previous sections that the cluster component is perturbed by the radio emitting plasma and is displaced from
the central peak. To evaluate the effect of these asymmetries on the
extracted profile we compare, in Figure \ref{ngc507_model_NEprofile}, the average ($0^\circ-360^\circ$) profile with the one extracted in the NE ($0^\circ-90^\circ$) quadrant, i.e. in the opposite
direction of the cluster component centroid and where the influence of radio
lobes is smaller. The comparison shows that the cluster component offset is
responsible for the flattening of the X-ray profile in the range $10"<r<30"$%
. For $r>80"$ the X-ray profile is almost unchanged since the displacement
is not relevant on large scales. We notice that this displacement affects
the total profile but not those of the individual components. The
consequences of this result will be discussed further in $\S$ \ref
{ngc507_dens}.

In $\S$ \ref{ngc507_brightness} we noticed that at large radii ($r>200"$)
there seems to be more X-ray emission on the North and East side of NGC 507
than on the South and West ones. To verify this fact we derived X-ray
surface brightness profiles in 2 regions: a North-East sector ($20^\circ<$%
P.A.$<110^\circ$) and a South-West one ($110^\circ<$P.A.$<200^\circ$), and
extracted counts in 30" annuli centered on the X-ray centroid. The resulting
profiles are shown in Figure \ref{ngc507_pies}. Past 200" the count rate in
the North-East pie is systematically higher than the one in the SW pie, out
to $\sim 450"$. This excess is significant at the $3\sigma$ level. For $%
r>500"$ instead, the counts in the two pies are consistent within the errors.

To check if the observed excess in the NE sector is preferentially located
within NGC 507 and NGC 499 we further divided the NE sector in two $%
90^{\circ }$ pies. We found that the behaviour of the radial profiles is
essentially the same in the two subsections in which the NE pie was divided,
thus indicating that the excess emission is distributed through all the NE
region.

\subsection{Density, Cooling Time and Mass Profiles}

\label{ngc507_dens}

\begin{figure}[]
\centerline{\psfig{figure=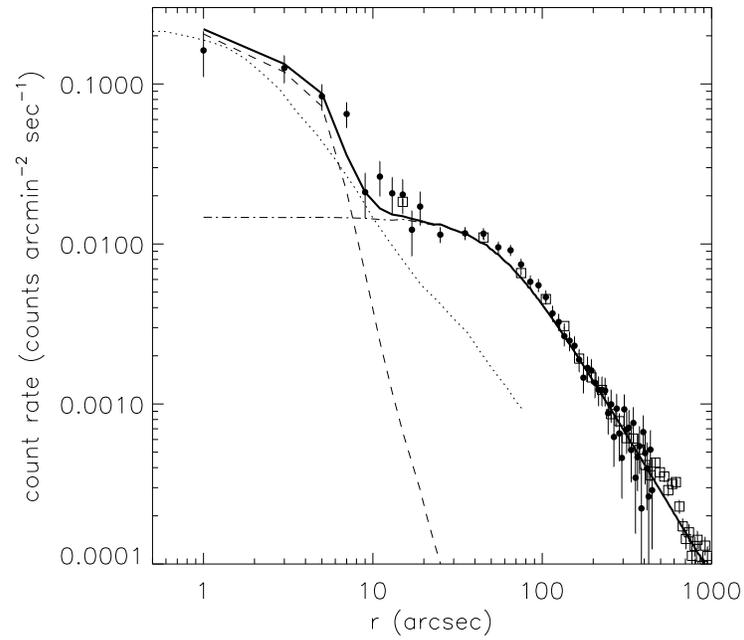,width=0.6\textwidth}}
\caption{Radial profile of the bidimensional HRI model (continuous line).
The central and extended components are represented respectively by the
dashed and dot-dashed lines. HRI (PSPC) counts are shown as filled circles
(empty squares). The dotted line shows the V band surface brightness profile
measured by \protect\cite{Gonzalez00}.}
\label{ngc507_model_profile}
\end{figure}

\begin{figure*}[]
\vspace{0.5cm} \centerline{\psfig{figure=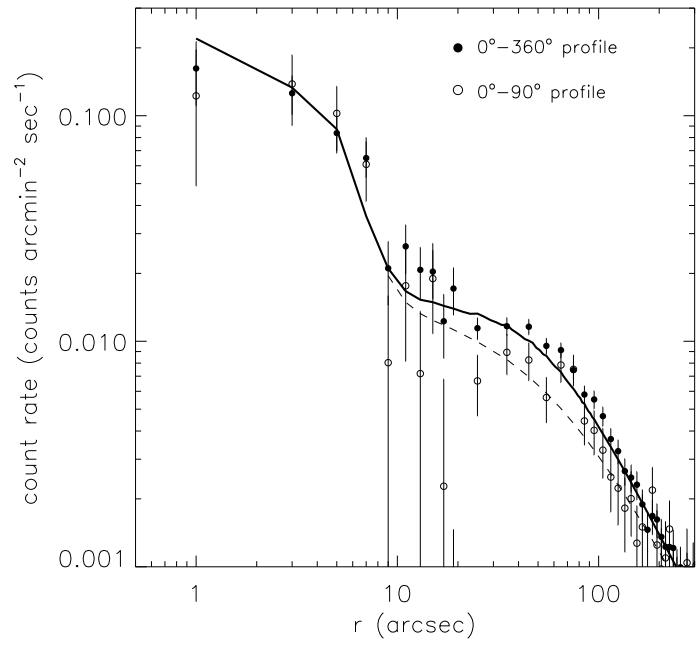,angle=0,width=0.6\textwidth}}
\caption{Comparison of the azimuthally averaged (filled circles,
continuous line) X-ray surface brightness profile (same as the one shown in Figure 
\ref{ngc507_model_profile}) with the one extracted in the NE quadrant (empty
circles, dashed line).}
\label{ngc507_model_NEprofile}
\end{figure*}

\begin{figure}[]
\vspace{0.5cm}
\centerline{\psfig{figure=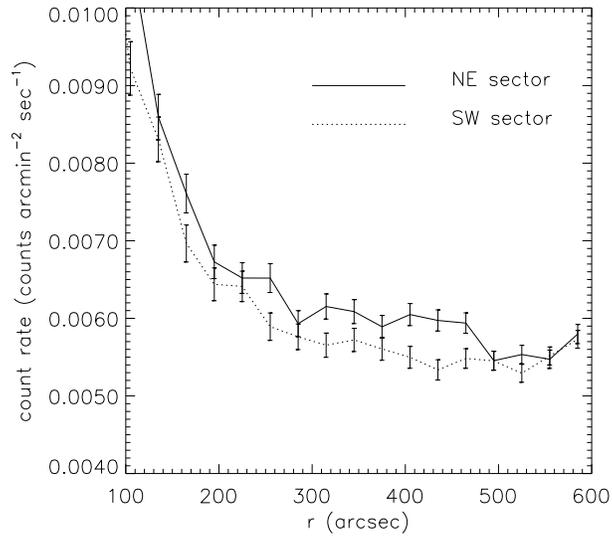,angle=0,width=0.5\textwidth}}
\caption{NE ($20^\circ<$P.A.$<110^\circ$) and SW ($110^\circ<$P.A.$<200^\circ$) radial surface brightness profile of the NGC 507 X-ray halo.}
\label{ngc507_pies}
\end{figure}

To derive the properties of the X-ray emitting gas we followed the
method described by \cite{kriss83}, deprojecting the measured emission in
concentric shells. We use the model profile shown in Figure 
\ref{ngc507_model_profile} , since the bidimensional model provides
a satisfactory representation of the mean X-ray emission, as shown in
$\S $ \ref{ngc507_brightness_prof}. This approach does not account for
the fluctuations in the X-ray surface brightness due to the interaction 
with the radio-emitting plasma, which are perturbing the X-ray halo within 
1 arcmin and for the displacement of the central and extended components. 
However, since we are mainly interested in obtaining the average gas properties,
we will use this approximation and point-out, in the following discussion,
the effects due to the lack of smoothness and spherical symmetry.

\begin{figure}[]
\vspace{0.5cm} \centerline{\psfig{figure=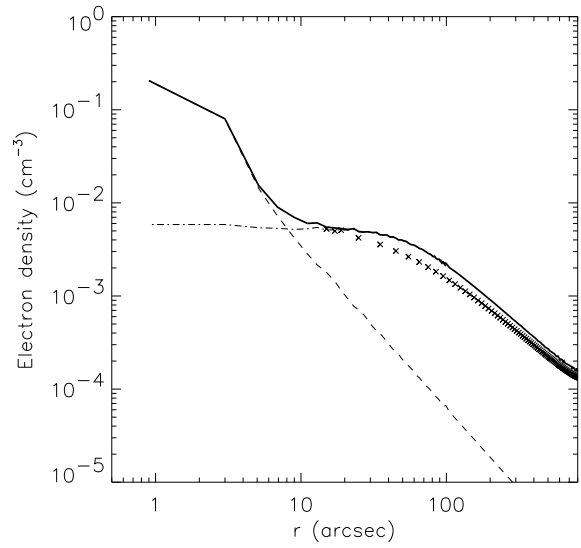,angle=0,width=0.5\textwidth}}
\caption{Deprojected density profile of the NGC 507 gaseous halo (continuous
line). The dashed and dot-dashed lines show, respectively, the contribution
of the central and extended component. Crosses represent the density obtained 
using the NE quadrant profile shown in Figure \ref{ngc507_model_NEprofile}.}
\label{ngc507_density_prof}
\end{figure}

We assume an isothermal temperature profile with $kT=1.1$ keV and the
cooling function given by \cite{Sar87}. Moreover, since the temperature
profile measured by \cite{Kim95} shows the presence of a central temperature
drop, we also used a lower temperature for the central component, obtaining
an almost identical result. The deprojected electron density is shown in
Figure \ref{ngc507_density_prof}.

\begin{figure}[]
\centerline{\psfig{figure=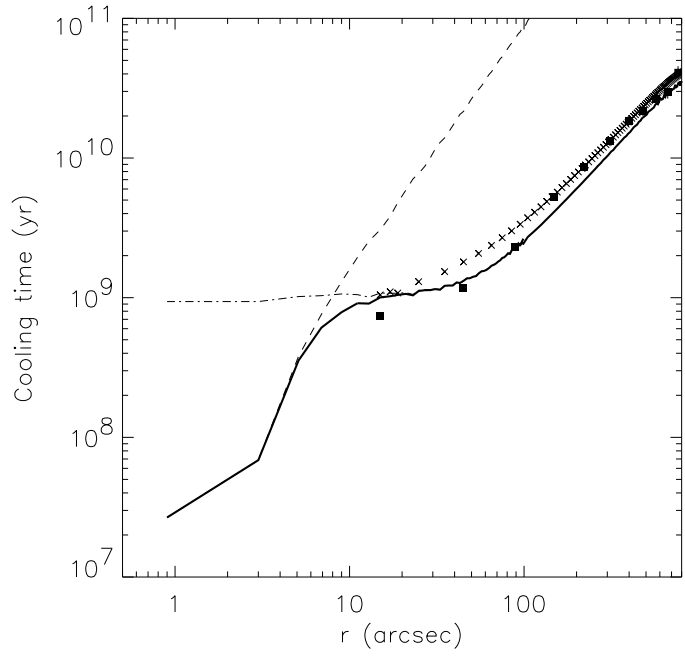,angle=0,width=0.6\textwidth}}
\caption{Cooling time profiles of NGC 507. Symbols have the same meaning
as in Figure \ref{ngc507_density_prof}. The cooling times derived by 
\protect\cite{Kim95} are shown as filled squares.}
\label{ngc507_CT}
\end{figure}

\begin{figure*}[]
\centerline{\psfig{figure=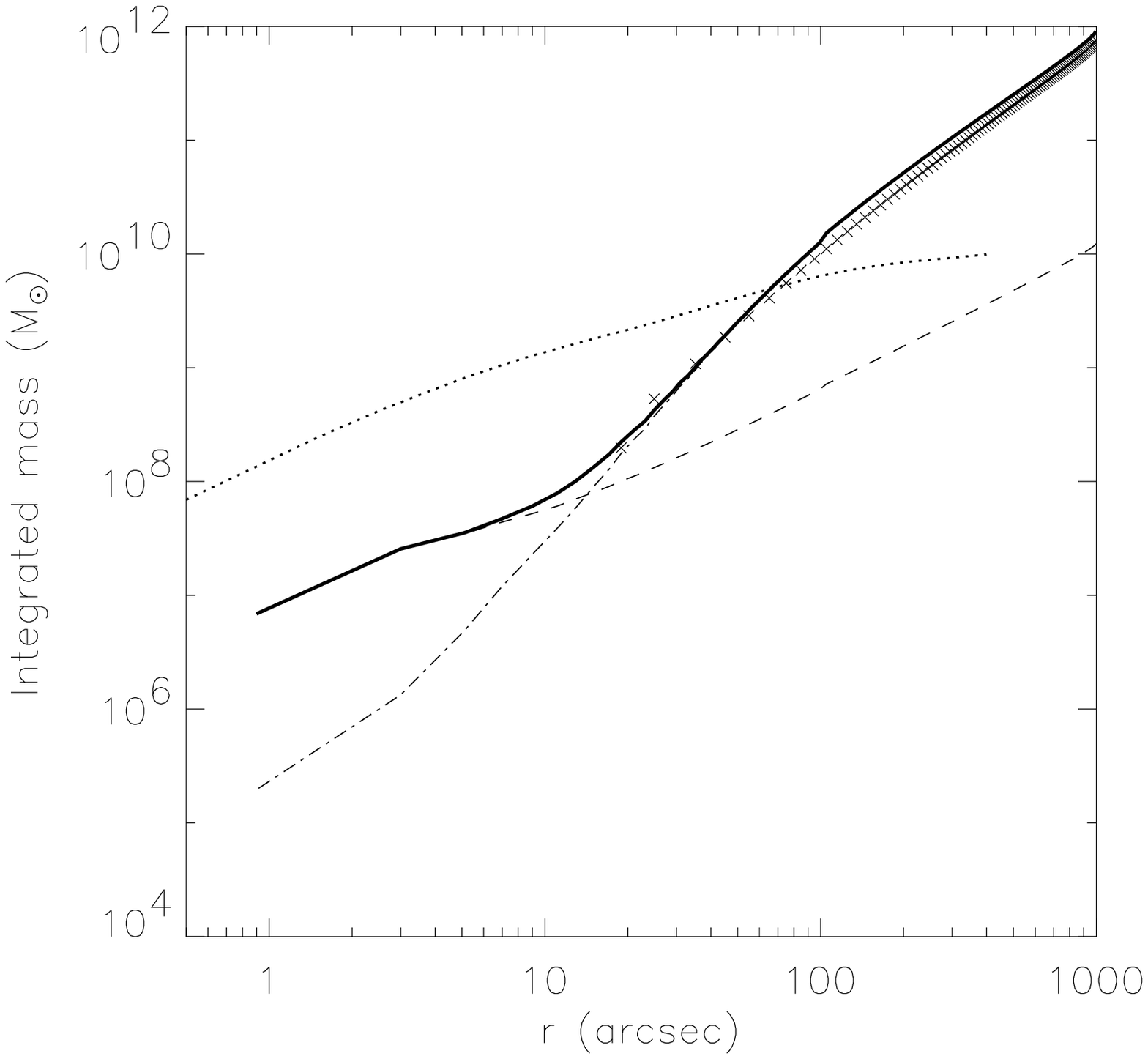,angle=0,width=0.5\textwidth}
\psfig{figure=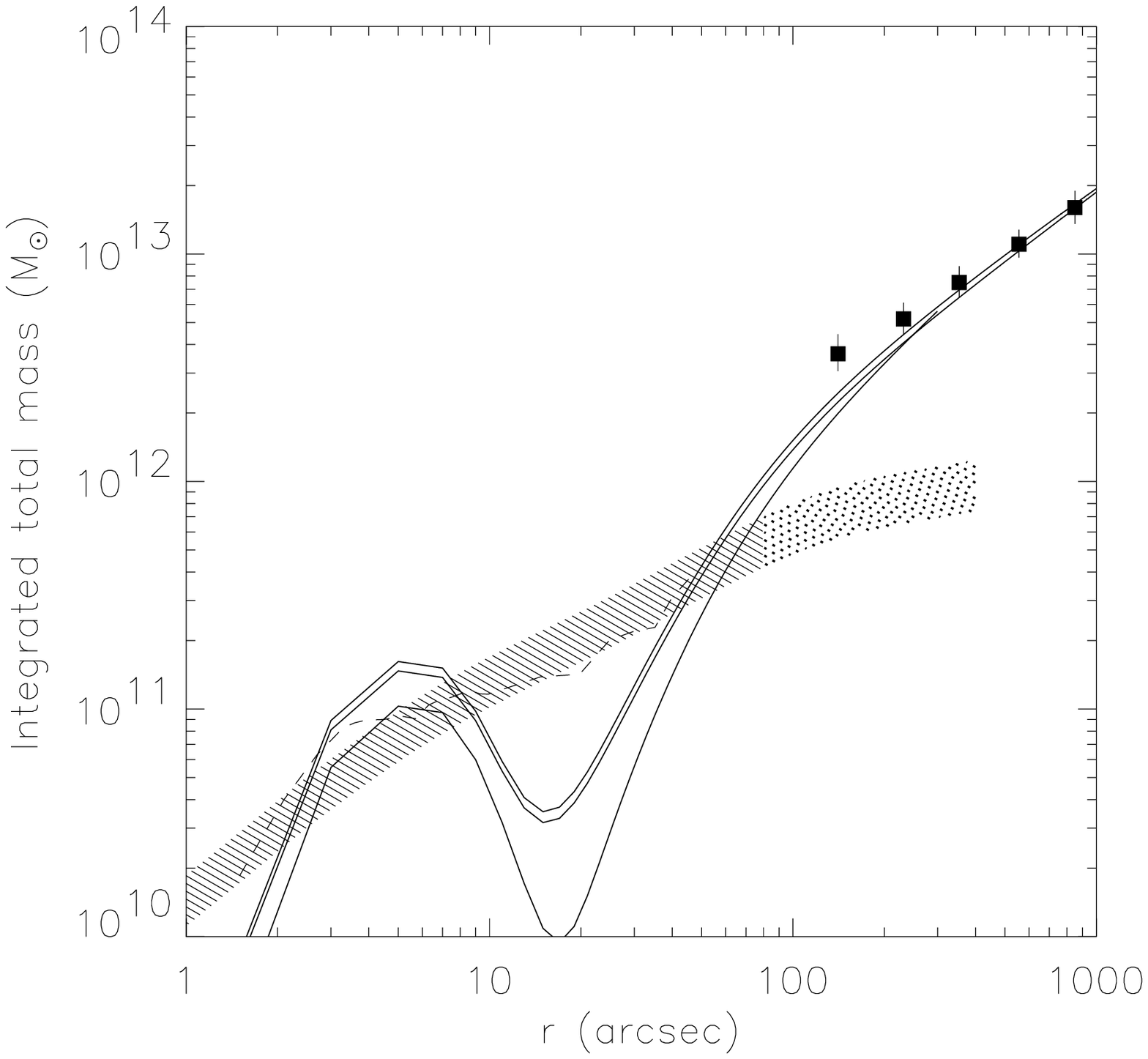,angle=0,width=0.5\textwidth}}
\caption{ {\it Left:} Integrated gaseous mass within $r$. 
Symbols are the same as in Figure \ref{ngc507_density_prof}. 
The dotted line represents the mass injected in the ISM by stellar mass losses. 
{\it Right:} Integrated total mass within $r$. The continuous
lines represent the gravitating mass estimates obtained assuming the
different temperature profiles described in the text. The total mass
estimate of \protect\cite{Kim95} is shown as filled squares. 
The dashed line shows the total mass  obtained using the surface brightness 
profile extracted in NE quadrant (see Figure \ref{ngc507_model_NEprofile}). 
The shaded (dotted) region shows the stellar mass contribution estimated (extrapolated)
from the optical profile measured by \protect\cite{Gonzalez00}, assuming a
M/L ratio ranging from 6 to 8 M$_\odot$/L$_\odot$.}
\label{ngc507_mass}
\end{figure*}

Using this density profile we calculated the cooling time $\tau_c\propto
nkT/n_e n_H \Lambda$ as a function of radius shown in
Figure \ref{ngc507_CT}, where $n_e$ is the electron density, $n_H=n_e/1.21$ 
\citep{Sar88} and $n=n_e+n_H$ are
the hydrogen and total density and $\Lambda$ is the cooling function. Our
result is in good agreement with the \cite{Kim95} estimate at large
galactocentric distances ($r>60"$) with minor differences due to their use
of a non-isothermal temperature profile. At smaller radii our HRI data are
able to sample the characteristics of the central component, whose cooling
time is more than one order of magnitude smaller than that of the extended
component. Both components, however, have cooling times smaller than the
Hubble time ($10^{10}$ yr) within the cooling radius $\simeq 250"$ (80 kpc).
Figure \ref{ngc507_mass} (left panel) shows the integrated gaseous mass as a function of radius.
The gaseous mass, calculated from the gas density profile assuming cosmic abundances \citep{Sar88}, yields $\simeq 10^{12}$ M$_\odot$ within 1000" (320 kpc). The contribution
of the central component, which dominates the inner 15" (5 kpc), is
estimated to be $\sim 10^{10}$ M$_\odot$ at $r=1000"$, representing $\sim
1\% $ of the total gaseous mass.
In Figure \ref{ngc507_mass} (left panel) we also plot the amount of mass
expected to be injected by stellar mass losses in $10^{10}$ yrs using a 
mass-loss rate of $0.0078\times 10^{-9}$ M$_\odot$/L$_{\odot,B}$ 
\citep{Athey02} and the optical profile shown in Figure 
\ref{ngc507_model_profile}.

The dependence of these results on the assumption of spherical symmetry 
was evaluated comparing the average gas properties to those derived using 
only the NE quadrant profile shown in Figure \ref{ngc507_model_NEprofile}. 
The main differences (see crosses in Figures \ref{ngc507_density_prof}, 
\ref{ngc507_CT} and \ref{ngc507_mass}, left panel) are seen in the 20-100 
arcsec region where, using the NE profile, the `shoulder' corresponding to 
the core radius of the extended component is smoother, thus decreasing the 
gas density and increasing the cooling time by $\sim 30-40\%$. At larger 
radii the gas properties are almost unchanged since the displacement is 
not relevant on large scales.

The total mass profile, shown in Figure \ref{ngc507_mass} (right panel),
was calculated using the equation \citep{Fabr80}: 
\begin{equation}  \label{mass}
M(<r)=-{\frac{rkT(r)}{\mu m_p G}}\left({\frac{\mbox{d}\log\rho_g}{\mbox{d}%
\log r}}+{\frac{\mbox{d}\log T}{\mathrm{d}\log r}}\right)
\end{equation}
where $M(<r)$ is the mass contained within $r$, $T$ is the gas temperature,
and $\mathrm{d}\log\rho_g/\mathrm{d}\log r$ and $\mathrm{d}\log T/\mathrm{d}%
\log r$ are respectively the logarithmic gradients of the gas density and
temperature.

We assumed three different temperature profiles: an isothermal profile at
1.1 keV, a linear fit $kT=-2.3\times 10^{-4}~r+1.21$ keV ($r$ is expressed
in arcsec) to account for the temperature decline after $\sim 150"$, and a
power law $kT=0.04~r^{0.5}+0.7$ keV to fit the central temperature drop.
Figure \ref{ngc507_mass} (right panel) shows that the total mass profiles
are weakly dependent on the assumed temperature profile, except in the inner
100" (32 kpc) where the central mass obtained accounting for the low central
temperature is $\sim 50\%$ smaller. The shaded region represents the total
stellar mass calculated from the surface brightness profile measured by \cite
{Gonzalez00} and assuming a M/L ratio ranging from 6 to 8 M$_\odot$/L$_\odot$.

For $r>50"$ our mass profile is in good agreement with \cite{Kim95} and
suggests the presence of dark matter, since it exceeds the stellar mass
estimate extrapolated from the optical profile measured by \cite{Gonzalez00}. 
For r$<50"$ the X-ray cumulative mass estimate yields an unphysical
result: the total mass estimate is smaller than the stellar mass 
and the profile has a negative gradient for $r>10"$ reaching a minimum at $\sim 20"$. This behaviour is due to the flattening of the density profile which produces a drop in the density derivative contained in equation (\ref{mass}). In fact, using the
profile extracted in the NE quadrant (dashed line in Figure \ref
{ngc507_mass}, right panel), the mass profile is almost in agreement
with the stellar profile. Thus the lack of azimuthal symmetry around the
central X-ray peak, produced by the offset between the central and extended
components and by the perturbed gas distribution, suggest that equation (\ref{mass}) is no longer valid on scales smaller than 50'' because the gas is not in hydrostatic equilibrium. The difficulties involved in deriving accurate mass profiles in the halo
core from X-ray measurements have been already noticed in previous works
\citep[see for instance][]{Brighenti97} and can be explained by the presence of non-gravitational effects
 such as those discussed in $\S$ \ref{ngc507_Xradio_discuss}\footnote{ Within a few arcsec from the main emission peak we must also consider the
influence of the PRF. Even if the bidimensional model accounts for
instrumental effects, an accurate deconvolution is very difficult.}.

\subsection{Discrete Sources}

\label{ngc507_sources}

\begin{figure}[p]
\centerline{\psfig{figure=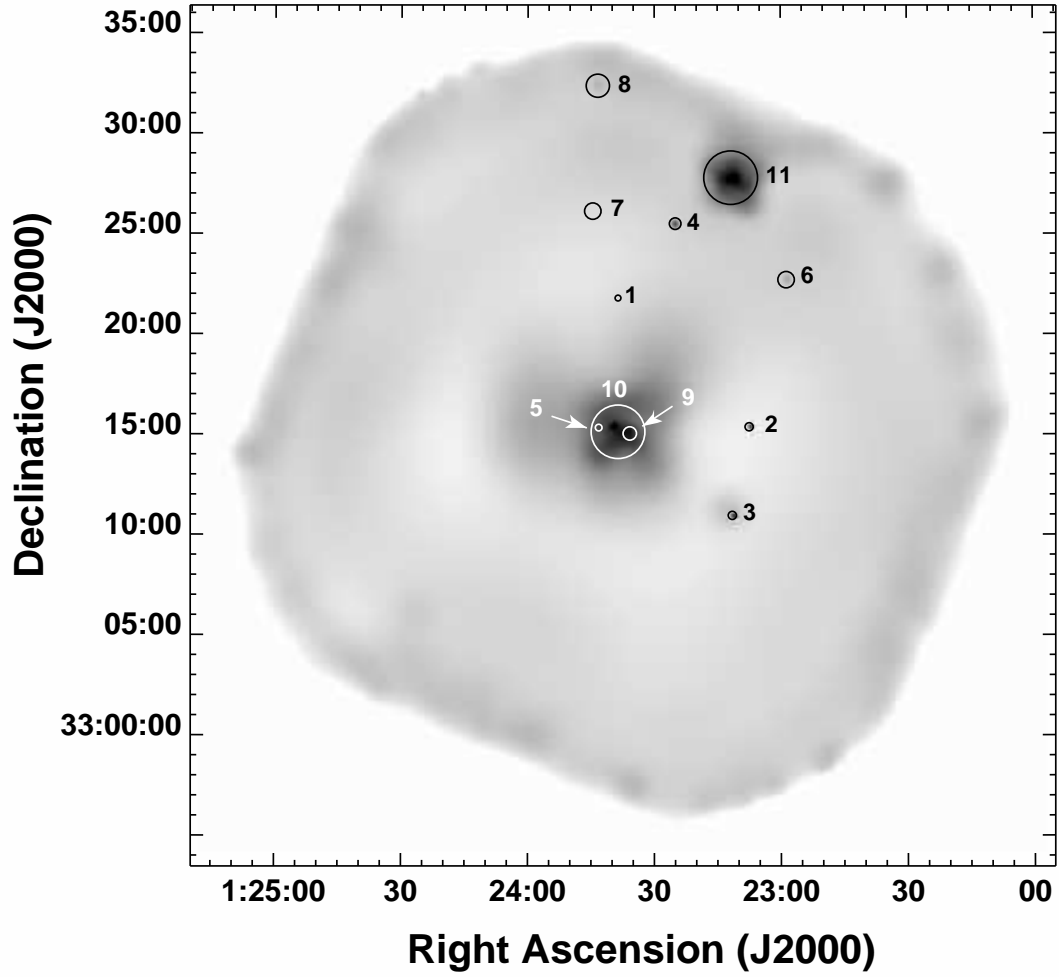,angle=0,width=0.9\textwidth}}
\caption{Discrete sources detected by the wavelets algorithm superimposed on
the adaptively smoothed HRI field. The circles show the region of maximum
S/N ratio for each source. The position of sources No.5 and 9 is shown in
greater detail in the left panel of Figure \ref{ngc507_centbox}.}
\label{ngc507_sources_fig}
\end{figure}

To study the population of discrete sources in the HRI field we used the
wavelets algorithm developed by \citet{Dam97a,Dam97b}. Using a S/N threshold
of 4.2 (corresponding to a contamination of 1 spurious source per field;
F.Damiani, private communication) the algorithm detected 11 sources shown in
Figure \ref{ngc507_sources_fig}. The properties of these sources, as
measured by the wavelets algorithm, are reported in Table \ref
{ngc507_source_tab}. Sources No.10 and 11 are missing from the table since
they are NGC 507 and NGC 499. We also notice that sources No.5 and 9
correspond to the secondary central peaks discussed in $\S$ \ref
{ngc507_brightness}.

To derive the X-ray fluxes we used the PIMMS software to calculate a
conversion factor between the HRI count rate and the incident flux. We
assumed a power law spectral model with photon index $1.96\pm0.11$ ($\mathit{%
f}_{HRI}^{pow.law}$ in Table \ref{ngc507_source_tab}) following the results
of \citet{Has93} for faint sources, and a thermal spectrum with $kT=0.52$
and abundance $Z\sim 0.2~Z_\odot$ ($\mathit{f}_{HRI}^{RS}$ in Table \ref
{ngc507_source_tab}) as found by \cite{Kim95}. Assuming a galactic
absorption of $5.3\times 10^{20}$ cm$^{-2}$ we obtain a conversion factor of
1 cps=$6.925\times 10^{-11}$ ergs s$^{-1}$ and 1 cps=$4.076\times 10^{-11}$
ergs s$^{-1}$ respectively in the 0.1-2.4 keV energy range.

In Table \ref{ngc507_source_tab2} we cross-correlate the sources detected in
the HRI field with those found by \citet{Kim95}. The fluxes that they derive
from PSPC data are reported in column $\mathit{f}_{PSPC}^{RS}$. We searched
for long-term variability of the matching sources comparing these fluxes
with those measured by the wavelets algorithm (Table \ref{ngc507_source_tab}%
). We mark as variable those sources whose flux varied by more than $3\sigma$
between the PSPC and HRI observations, i.e. between 1993 and 1995. We found
that only source No.3 show significant variability in this interval. To
check if the different procedures, extraction radii and background
subtraction, significantly affect the PSPC count rate determination, we
extracted from the GALPIPE database \citep[ and references therein]{mack96},
which has produced a list of all sources present in the ROSAT PSPC archive%
\footnote{%
An online version of the GALPIPE database and documentation can be found at
the D.I.A.N.A. homepage at the Palermo Observatory:
http://dbms.astropa.unipa.it/}, the count rates measured by the wavelets
algorithm on the second of the PSPC observations (RP600254a01) used by %
\citet{Kim95}. The results are in good agreement with the \citet{Kim95}
determinations, except for source No.4. In this case however the proximity
to the NGC 499 halo and the large PSPC PRF may compromise the wavelets
accuracy. We must notice that sources No.5 and 9, which are likely to be
produced by the interaction of the radio lobes with the surrounding ISM (see 
$\S$ \ref{ngc507_Xradio_discuss}), have no PSPC counterpart since they are
undetected due to the lower spatial resolution of this instrument.

The optical DSS image of the NGC 507/499 region was visually inspected to
find optical counterparts of the X-ray sources within a $2\sigma$ radius
(Table \ref{ngc507_source_tab}) from the X-ray centroid. The results are
summarised in column ``Opt.id.'' of Table \ref{ngc507_source_tab2}. 
We also cross-correlated our source list with the catalogs contained in the NED
database, finding that source No.7 is likely to be the X-ray counterpart
of the radio source NVSS J012345+332554.

Many of the sources detected in the PSPC data by \citet{Kim95} are not
detected in the HRI field. We derived upper limits on the HRI flux to find
whether this result is due to the lower HRI S/N ratio or by the temporal
variability of the sources. For all PSPC sources not detected in the HRI
field, we extracted count rates using the radii reported in Table 2 of %
\citet{Kim95} rescaled by a factor 1/3 to take into account the difference
between the HRI and PSPC PRF. The background was measured locally in an
annulus of radius twice the one used for the source. The resulting fluxes
(obtained with the same conversion factors discussed above) and $3\sigma$
upper limits, are shown in Table \ref{ngc507_source_tab3}. The comparison
with the PSPC fluxes shows that all sources except one are below the HRI
sensitivity. Source No.6 is the only one which must be significantly
variable since the expected flux is higher than the $3\sigma$ upper limit
indicated by the HRI data.

The implications of these results on the nature of the discrete sources
found in the NGC 507/499 field will be discussed in $\S$ \ref
{ngc507_sources_discuss}.

\subsection{The \textit{Chandra} data}
\label{ngc507_chandra} 
\textit{Chandra} observed NGC 507 on 2000 October 11
(ObsID 317) with the ACIS-S detector for $\sim 28$ ks. The \textit{Chandra}
field of view is smaller than the one of the Rosat HRI and is not
azimuthally symmetric due to the position of the ACIS ccd's on the focal
plane. Thus the \textit{Chandra} data are not as good as the Rosat one's to
relate the core region to the large scale structure of the X-ray halo;
however the higher angular resolution allows a more detailed study of the
NGC 507 core. In this work we use the \textit{Chandra} data to confirm our
HRI results and to study the properties of the central peak; a complete
analysis of the \textit{Chandra} data is already in progress %
\citep{Forman01a,Forman01b}.

\begin{figure}[t]


\centerline{\psfig{figure=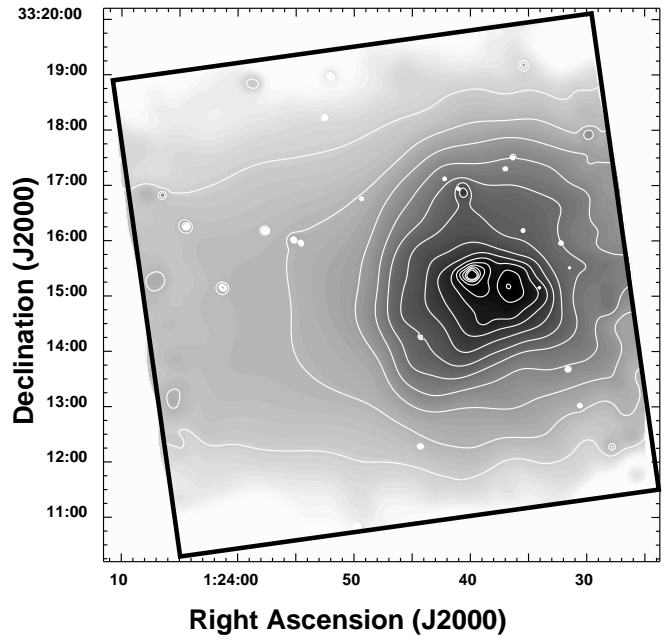,angle=0,width=0.58\textwidth}}
\caption{\textit{Chandra} ACIS-S adaptively smoothed image of NGC 507. 
The black box shows the edges of the ACIS-S3 chip. Contours are spaced by 
a factor 1.2 with the lower one at $2.4\times 10^{-2}$ photons arcmin$^{-2}$ 
s$^{-1}$. A higher quality color version of this figure is available in the electronic 
edition.}

\label{ngc507_ACIS}
\end{figure}

The adaptively smoothed and exposure corrected image of the ACIS-S3 chip is
shown in Figure \ref{ngc507_ACIS}. The \textit{Chandra} data confirm the
complex morphology shown by the Rosat HRI: the galaxy has an extended halo
with two X-ray peaks and a small gas tongue extending in the South-East
direction, as discussed in $\S$ \ref{ngc507_brightness}. The presence of the
Northern `tail' seen in Figure \ref{ngc507_csmooth} is less evident; however
the surface brightness on the Northern side extends in the North-West
direction and its gradient is shallower than toward the South. This is also
in agreement with the presence of sharp Southern and Eastern edges in the
gas distribution reported by \cite{Forman01b}, even though in the adaptively
smoothed image they are not as evident as in the raw data. As discussed in $\S$ \ref{ngc507_brightness_prof}, the surface brightness profile of the central component \citep[see][]{Forman01a} demonstrates that the central X-ray peak is extended and not due to the presence of a nuclear source, thus solving any ambiguity present in the HRI data.

Figure \ref{ngc507_ACIS} confirms the presence of a clump of hot gas in the
core of the companion galaxy NGC 508. The count rate measured with 
\textit{Chandra} ($3.2\pm0.6 \times 10^{-3}$ cnts s$^{-1}$) is consistent 
with the HRI detection and yields an X-ray luminosity\footnote{The error 
includes the uncertainty on the assumed spectral model, ranging from a 
thermal plasma with variable metallicity to a 5 keV Bremsstrahlung spectrum} 
$L_X\sim 1.5\pm 0.5\times 10^{40}$ erg s$^{-1}$. In the L$_X$-L$_B$ diagram 
this places NGC 508 (optical luminosity $L_B=1.7\times 10^{10} L_\odot$ 
\citep{RC3}) in the range occupied by compact early-type galaxies, whose 
emission includes a large contribution from X-ray binaries.

\begin{figure*}[t]
\centerline{\psfig{figure=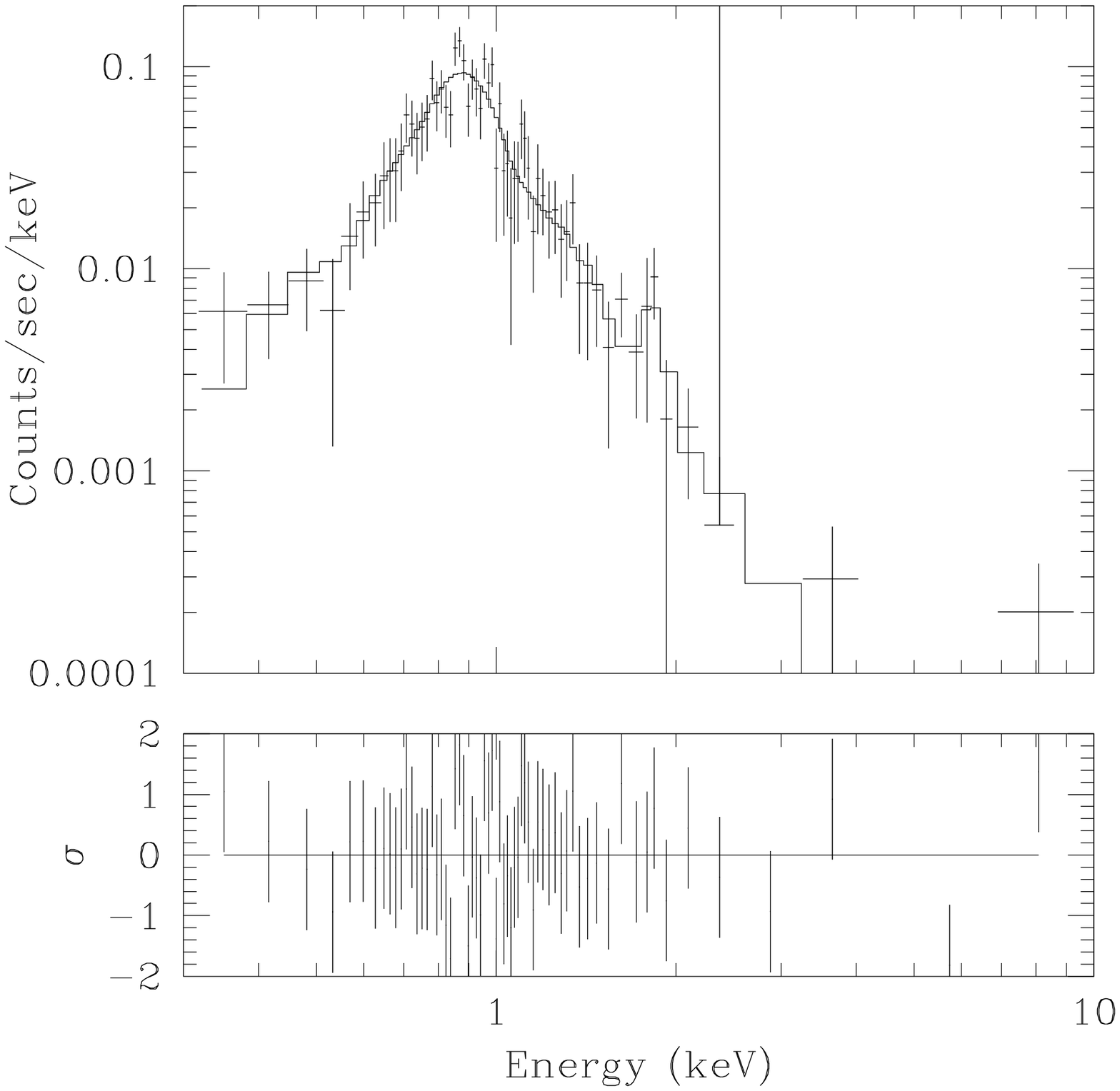,angle=0,width=0.6\textwidth}
\psfig{figure=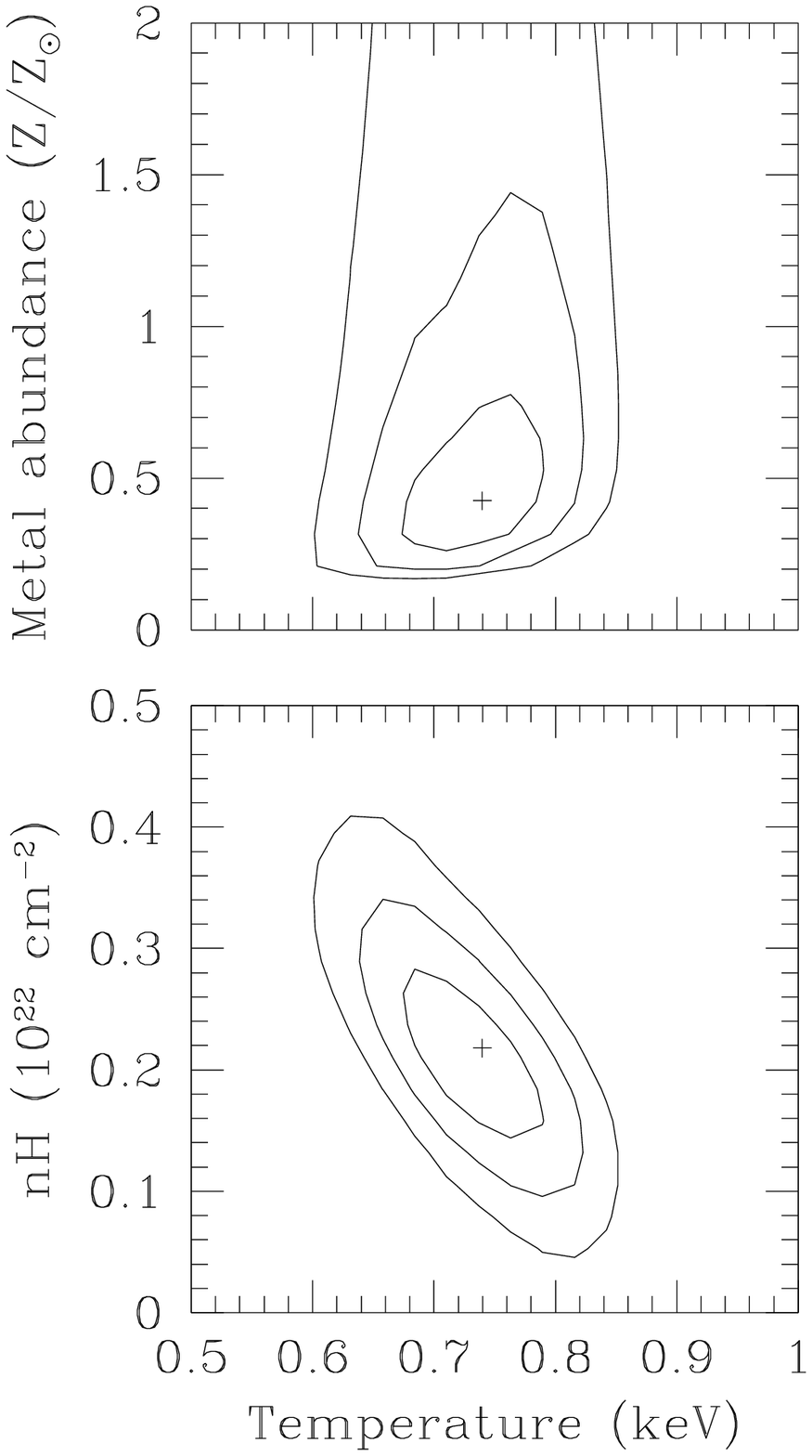,angle=0,width=0.56\textwidth}}
\caption{\textbf{Left}: Best-fit MekaL spectral model of the central 10
arcsec region. Residuals are shown in the bottom panel. \textbf{Right}: 1, 2
and 3$\protect\sigma$ confidence levels for the best-fit MekaL model.}
\label{ngc507_mekal_fit}
\end{figure*}

The preliminary spectral analysis performed by \cite{Forman01b} shows that
the mean halo temperature is close to 1 keV, in agreement with the
previously published ROSAT PSPC \citep{Kim95} and ASCA data \citep{Matsumo97}%
. Taking advantage of the \textit{Chandra} angular resolution, we further
analysed the spectral properties of the central component. We extracted the
0.3-10 keV spectrum from a 10 arcsec circle centered on the central peak,
using the surrounding 10-20 arcsec annulus to extract the background. The
resulting counts were rebinned so to have at least 15 counts per bin and
fitted with the CIAO \textit{Sherpa} package. We tried both a single
temperature Raymond-Smith \citep[ and additions]{RS77} and a MekaL %
\citep{Mewe85,Mewe86,Kaastra92,Liedahl95} spectral model obtaining the
results reported in Table \ref{ngc507_spectral_tab}. The mean gas
temperature is $kT\sim 0.8$ keV for the Raymond-Smith model and $\sim 0.7$
keV for the MekaL one (Figure \ref{ngc507_mekal_fit}), with the MekaL model
yielding slightly better fits than the Raymond-Smith one. The metal
abundances are subsolar but compatible with solar within $3\sigma$. The
absorbing columns (nH) are 2-3 times higher than the galactic one ($%
5.3\times 10^{20}$ cm$^{-2}$, \citealp{}) even though with large
uncertainties. Fixing the abundance to the solar one yields slightly higher
temperatures and lower absorbing columns, while fixing the nH to galactic
values produces metal abundances higher than solar. However, even though the
small statistics produces large errors, the plasma temperature is quite well
constrained as shown by the confidence contours in Figure \ref
{ngc507_mekal_fit} (right panel).

We further tried to reduce the extraction region to a 5 arcsec circle, to
find out if there are gradients in the spectral parameters of the central
component. The results of these fits are shown in the right section of Table 
\ref{ngc507_spectral_tab}. The plasma temperatures are slightly lower than
those extracted within 10 arcsec for all models while the value of the other
parameters are very uncertain due to the small statistics. However the
results of the fits suggest that the metallicity may be higher in the galaxy
center.

We are aware that to correctly estimate the hot plasma properties we should
perform a more complex analysis, e.g. a full spectral deprojection of the
different halo components, since there is a temperature gradient in the NGC
507 halo. Several authors have shown that an incorrect modelling of the X-ray
emission (e.g. single instead of multi-temperature models) leads to
systematic errors in the gas parameter estimates \citep{Buo99,Mat00}.
However in our case the small number of events does not allow a more
complex analysis of the nuclear region. We tried to add the contribution of
unresolved discrete sources in the form of a 5 keV Bremsstrahlung or a
power-law component but the fit yields a negligible contribution and,
however, does not significantly affect the best fit temperature. The
best-fit temperature does not change even if we restrict the fit to the
0.3-4 keV range, where the plasma emission is the dominant contribution. In
conclusion we believe that the central component temperature is well
constrained to be $\sim 0.7$ keV.

\section{DISCUSSION}

\subsection{Origin of the Central Component}

\label{ngc507_central_discuss} The surface brightness distribution shown in
Figure \ref{ngc507_centbox} revealed the presence of a bright X-ray emission
peak centered on the optical galaxy. The surface brightness profiles (Figure 
\ref{ngc507_profile} and \ref{ngc507_model_profile}) show that the peak
dominates the X-ray emission within the central 10 arcsec (3 kpc). The HRI
data alone do not allow to distinguish between a pointlike or extended
nature of the central peak, since the comparison of the X-ray profile with
the HRI PRF (Figure \ref{ngc507_profile}) and the small core radius of the
central component obtained from the bidimensional fit (Table \ref
{ngc507_fit_tab}) are marginally consistent with the presence of a nuclear
X-ray source. However the \textit{Chandra} surface brightness profile
obtained by \cite{Forman01a} demonstrates the extended nature of this
component.

The comparison between the optical and X-ray profile (Figure \ref
{ngc507_model_profile}) seems to rule out the possibility that the central
peak is due to the unresolved emission of galactic X-ray binaries. In this
case the X-ray profile of the central component would be close to the
optical one, since the X-ray source population should follow the stellar
distribution. The higher \textit{Chandra} angular resolution and the
spectral fit of the central component ($\S$ \ref{ngc507_chandra}) further
showed that the emission is mainly due to hot plasma and that the
contribution of unresolved point sources is small, supporting the gaseous
nature of the central peak.

The presence of a central X-ray excess in galaxies and clusters has been
usually associated to the presence of a cooling-flow. \cite{Kim95} noticed
that significant mass deposition should be present within the cooling radius
since the cooling times of the gaseous halo are smaller than the Hubble time
(Figure \ref{ngc507_CT}). Their estimate, rescaled to our adopted distance,
yields a mass inflow rate of $\sim 20$ M$_\odot$ yr$^{-1}$. If the central
X-ray peak was due to the gas deposited by the cooling flow, we would
expect to find $\sim 2\times 10^{11}$ M$_\odot$ within 250 arcsec from the
galaxy center while the gaseous mass of the central component is only $\sim
2\times 10^{9}$ (Figure \ref{ngc507_mass}, left panel) within the same
radius. This suggests that the cooling flow is inhomogeneous \citep{Kim95}
so that a large amount of cooling gas is deposited at large radii and 
does not reach the halo center. However, if we assume an inhomogeneous flow where 
$\dot{M}\propto r$ \citep{Fabian94} and extrapolate the mass deposition
down to $r=10"$, where the contribution of the central component dominates,
we still obtain a deposited mass of $\sim 10^{10}$ M$_\odot$ while we observe less
than $10^{8}$ M$_\odot$. This result holds also if we take into account all
the uncertainties included in the calculation such as the gas filling factor
and the age of the cooling flow.
Moreover we must consider that stellar mass losses are expected to
have injected an additional $10^9$ M$_\odot$ of gas in the central 10" 
(see Figure \ref{ngc507_mass}, left panel), further increasing the disagreement.

Thus, if the cooling flow model is correct, most of the deposited 
gas must have cooled out of the X-ray emitting phase. 
The fate of the cooling gas has been largely debated in
literature \cite[see for instance][]{Fab89,Fabian94,rang95,Fabian01} and is not clear yet.
In NGC 507 the column density profile derived by \cite{Kim95} is
consistent with the galactic value and shows that this mass is not in the form of
cold absorbing gas. One possibility is suggested by \cite{Buo00b},
who found that a large amount of
warm ($kT\sim 10^{5-6}$ keV) ionised gas may be present in the halo of NGC 507,
consistent with the one expected to be deposited by the cooling flow.
In this case we must address the question of what prevents this gas 
to rapidly cool to lower temperatures.
Alternatively a significant amount of gas may have been removed by star formation.
This hypothesis however, is challenged by the recent failure to observe the
cooling gas in the center of galaxies and clusters supposed to host large cooling 
flows \citep{Matsu02,Pet01,Kaa01,Tam01}.

The problems mentioned above have led many authors to stress the 
failure of the standard cooling flow models
and the need to re-analyse the nature of the central X-ray excess
\citep[e.g.][]{ikebe99,Maki01}.
The multi-component bidimensional fit of the X-ray surface brightness
distribution performed in the present work for NGC 507 and by \cite{paolillo02}
 for NGC 1399 has shown that the central component is centered
on the optical galaxy even though the outer halo is usually asymmetric and
displaced several kpc (7 kpc for NGC 507, 5 kpc for NGC 1399) from the
optical centroid. This suggests that the distribution of the denser central
gas is related to the underlying stellar distribution.

\begin{figure}[t]
\centerline{\psfig{figure=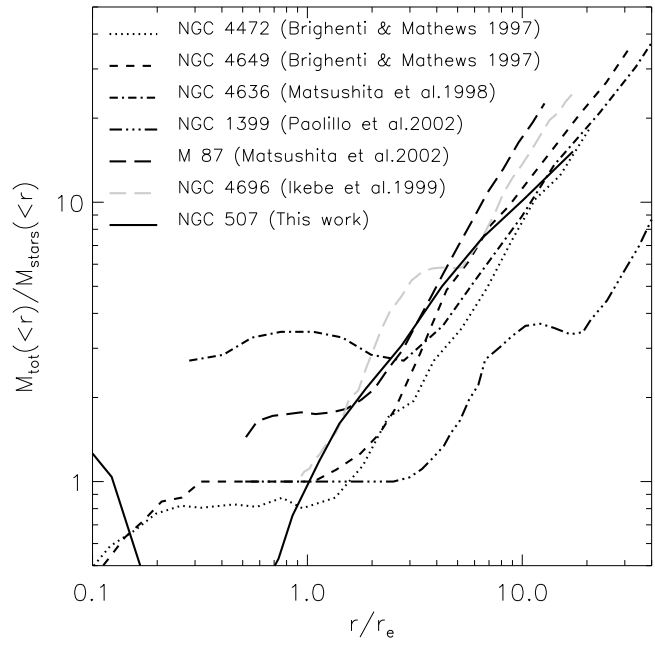,angle=0,width=0.6\textwidth}}
\caption{Total over stellar mass profiles derived from literature. Total
mass estimates are derived from X-ray measurements while the stellar
contribution is obtained from optical and dynamical estimates.}
\label{mass_prof}
\end{figure}

Following \cite{Maki01} and \cite{Matsu02} we compared the total
over stellar mass profiles of several X-ray bright Early-Type galaxies
extracted from literature. Figure \ref{mass_prof} shows that the X-ray
halos are characterised by two different dynamical ranges: a central region,
within 2-3 $r_{e}$, where the total over stellar mass is almost constant
with $1<M_{Tot}/M_{\star }<3$, and an outer region where the 
$M_{Tot}/M_{\star }$ ratio increases steeply indicating the presence of
massive dark halos. The flat central region correspond to the one dominated
by the X-ray excess. The $M_{Tot}/M_{\star }$ ratio close to 1 indicate that
the dark matter contribution in the galaxy core is usually small; however
even when $M_{Tot}/M_{\star }$ is higher, the constant value for $r<3r_{e}$
suggests that the total mass is still related to the underlying stellar
distribution. The correlation of the `knee' in the $M_{Tot}/M_{\star }$
curves with the effective radius further supports the association of the
central X-ray peak with the optical galaxy. In the case of NGC 507 this
trend is not as evident as for the other galaxies of the sample for the
lack of hydrostatic equilibrium in the halo core discussed in $\S$ \ref{ngc507_dens}. However we notice that for $r> r_{e}$, where $r_{e}$ is reported in Table \ref{ngc507observations}) 
the $M_{Tot}/M_\star$ ratio is very similar to those of the other galaxies.

A second point which was usually invoked as a proof of the cooling flow
scenario was the temperature drop in the halo center. However \cite
{Brighenti97} argue that the temperature drop within 4 effective radii ($r_e$)
observed in many galaxies, including NGC 507 and NGC 1399, is not a
natural result of cooling flow models.
In a recent study of a sample of early-type galaxies, \cite{Matsu01}
explained the properties of X-ray compact galaxies as due to the gas
injected into the ISM by stellar mass losses. The X-ray luminosities and
temperatures are compatible with the expectations obtained considering
stellar motions as the main source of heat, leading to a mean value of $\beta_{spec}={\frac{\mu m_p\sigma^2_r}{kT_g}}\simeq 1$. X-ray extended
galaxies, instead, have a mean halo temperature which exceeds the value
expected if the gaseous component is in equilibrium with the stellar
population ($\beta_{spec}\simeq 0.5)$. This difference between compact and
extended galaxies is reduced if we restrict the analysis to the region
within $\sim 1~r_e$ from the galaxy center, due to the temperature gradient
observed in most X-ray extended galaxies of the sample. NGC 1399 has the $%
\beta_{spec}$ value closest to unity ($\beta_{spec}\simeq 0.9$) among those
included in the \cite{Matsu01} sample. This fact is easily explained
considering the analysis performed by \cite{paolillo02} which shows that the
central component dominates the X-ray emission within $1.5 r_e$.

In the case of NGC 507 (which is not in the \cite{Matsu01} sample), if the
gas is in equilibrium with the stellar component, we expect to find a
central temperature close to 0.7 keV (assuming $\sigma=330$ km s$^{-1}$, 
\cite{prugniel96}), which is in excellent agreement with the spectral
determinations based on the \textit{Chandra} data ($\S$ \ref{ngc507_chandra}%
). The lower central temperatures would thus be explained by kinematical
heating from stellar mass losses rather than by the cooling flow model.
In this scenario it is easy to explain why X-ray extended galaxies have $%
\beta_{spec}<1$ since to correctly sample the galaxy core we need an angular
resolution high enough to isolate the temperature of the central component
from the one of the outer halo, which is usually higher.

\begin{figure}[t]
\centerline{\psfig{figure=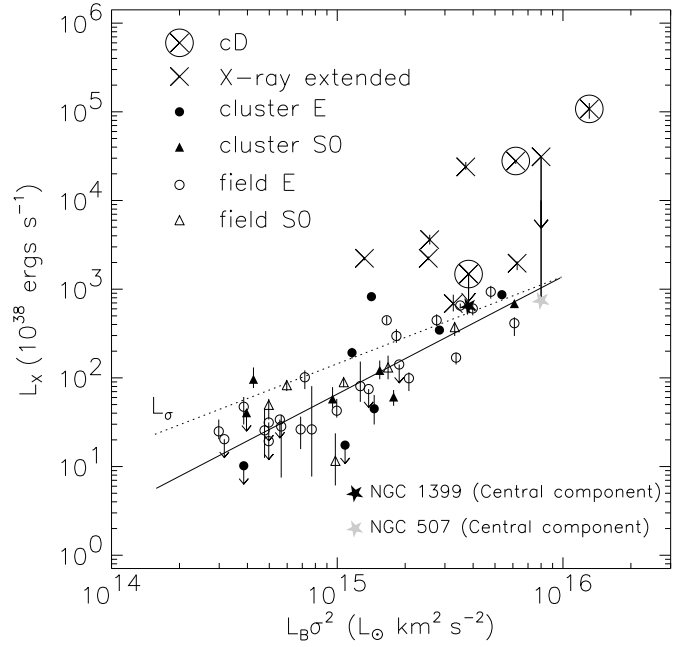,angle=0,width=0.6\textwidth}}
\caption{$L_X(<4r_e)$ vs $L_B\protect\sigma^2$ for a sample of Early-Type
galaxies. The Figure is identical to Figure 4 of \protect\cite{Matsu01}
except for the addition of NGC 507. The stars represent the drop in X-ray
luminosity for NGC 507 and NGC 1399 considering only the contribution of the
central component. The continuous line is the best fit to X-ray compact galaxies while the dotted line represents the
kinetic heating rate by stellar mass loss (see \citealp{Matsu01} for details)}.
\label{matsushita}
\end{figure}

To further confirm the association of the central component with the stellar
population we computed the X-ray luminosity, within $4 r_e$, from the
central component of NGC 507 and NGC 1399 and plotted the result (Figure \ref
{matsushita}) on the $L_X(<4r_e)$ vs $L_B\sigma^2$ diagram shown in Figure 4
of \cite{Matsu01}. The Figure shows that, removing the contribution of the external 
halo, the luminosity of the central component falls in the range occupied by
X-ray compact galaxies whose properties, according to \cite{Matsu01}, are well explained
by kinematical heating of the gas supplied by stellar mass loss.

In conclusion the characteristics of the central X-ray peak analysed in this
work for NGC 507 and in \cite{paolillo02} for NGC 1399 are compatible with
the scenario where the ISM in the galaxy core is produced by stellar mass
losses and is in equilibrium with the stellar component. Both gas and stars
feel the same potential, mainly due to stellar mass, since the amount of
dark matter is small. The stellar origin of the gas is further supported by
the solar metallicity found by \cite{Buo00a} in the center of several
galaxies and groups. 

The scenario that is emerging about the nature of the
X-ray excess in galaxies solves several problems (origin of the central
excess, lower central temperatures, subsolar abundances) largely debated in
literature in the last 10 years; however it also reveals our still poor
understanding of the cooling mechanism: if we assume that the 
gas remains in the hot phase, the X-ray data are inconsistent
with the deposition rates predicted not only by homogeneous models (a well
established fact) but by the inhomogeneous scenario as
well. Thus either the hot gas must cool out of the X-ray emitting 
phase, in which case {\it Chandra} and XMM observations are showing that we 
are still far from understanding the details of the cooling process and the 
fate of the cooling gas, or there must be some mechanism which prevents the 
hot gas to flow inward. The latter explanation may hold for NGC 507 as we will 
discuss in $\S$ \ref{ngc507_Xradio_discuss}.

\subsection{Large-scale Halo structure}

\label{large-scale} The large-scale properties of the X-ray halo are very
different from those of the halo core discussed in the previous section. We
have shown that the HRI data reveal ($\S$ \ref{ngc507_brightness}) the
presence of X-ray emission out to 500'' (160 kpc) from the galaxy center
(Figure \ref{ngc507_pies}) and more extended toward NGC 499. The mass
profiles (Figure \ref{ngc507_mass}) show that for $r>80"$ (25 kpc) the total
mass greatly exceeds the visible one. While for some galaxies %
\citep[e.g.][]{paolillo02} it can be argued that the hydrostatic equilibrium
assumption (on which the total mass estimates are based) is not valid due to
interactions with the surrounding environment, the good agreement of the NGC
507 $M_{Tot}/M_\star$ ratio with those derived from literature (Figure 
\ref{mass_prof}) shows that the prevalence of dark matter outside a few
effective radii is a common feature of many early-type galaxies.

If the hot gas distribution traces the gravitational potential this dark
matter is distributed preferentially on group scales rather than being
associated with the dominant galaxy of the cluster. In fact the study of the
X-ray surface brightness distribution showed that the optical galaxy is
displaced with respect to the gaseous halo. This result is confirmed
independently by the bidimensional fit, which shows that the extended
component centroid is located $\sim 22"$ (7 kpc) South-West of the optical
galaxy (Figure \ref{ngc507_model}, left panel), and by the wavelets
algorithm, which centered source No.10 (associated with the NGC 507 halo)
South-West of source No.9 (coincident with the central peak, see left panel
of Figure \ref{ngc507_centbox}). A very similar situation, where the central
X-ray peak is centered on the optical galaxy while the external halo is
displaced from the optical distribution, was found by \cite{paolillo02} for
NGC 1399, thus supporting the idea that the large-scale distribution traces
the gravitational potential of the galaxy group. In the case of NGC 507 the
displacement toward NGC 499 seems compatible with the distribution of
galaxies within the NGC 507 group studied by \cite{Barton98}.

The comparison of the halo temperature outside a few effective radii and the
stellar velocity dispersion in several early-type galaxies yields values of $\beta_{spec}\simeq 0.5$ \citep{Matsumo97,Brown98,Brown00,Matsu01} thus
supporting the need for additional heat to the one provided by stellar mass
losses alone. Sources of heat such as a central active nucleus or supernovae
explosions, seem not enough to account for the observational data %
\citep[e.g.][]{Matsumo97} and would however affect in first place the halo
center. The alternative explanation proposed by \cite{Maki01} and \cite
{Matsu01} is that X-ray extended galaxies are at the bottom of large
potential structures corresponding to galaxy groups, which provide the
additional heat by gravitational infall. This scenario is consistent with
our analysis which shows that NGC 507 and most X-ray bright early-type
galaxies, lies at the center of extended dark halos.

The displacement of NGC 507 with respect to the large-scale gas distribution
may be produced by the motion of the stellar body within the extended
gaseous halo. The central component would then be centered on the optical
galaxy since it is due to gas of stellar origin in equilibrium with the
stellar component ($\S$ \ref{ngc507_central_discuss}). The tails seen in the
adaptively smoothed images (Figure \ref{ngc507_csmooth} and Figure \ref
{ngc507_ACIS}) may represent cooling wakes produced by such motion, similar
those found by \cite{Dav94} or \cite{Merrifield98}. \cite{Forman01b}
suggested a similar origin for the sharp Southern and Eastern edges in the
gas distribution. The smaller North-Western tail departing from the central
component found in the surface brightness map (Figure \ref{ngc507_centbox})
may be produced by such motion as well, even though we can not exclude that
it simply represents a layer of gas compressed by the radio plasma on the
edge of the Western radio lobe.

Alternatively tidal interactions between NGC 507 and NGC 499 may be
responsible for the asymmetries in the hot gas distribution. Such
interactions may produce large density fluctuations and the formation of
filamentary structures \citep{D'Ercole00}. The presence of `clumps' of
cooler gas in the halo, suggested by \cite{Kim95} (see $\S$ \ref
{ngc507_sources_discuss}) may support such scenario.

These considerations do not necessarily exclude each other since a
combination of different effects is likely to be present in high density
environments.

\subsection{X-ray/Radio Correlations}

\label{ngc507_Xradio_discuss}

We showed in $\S$ \ref{ngc507_brightness} that the central NGC 507 halo has
a very complex morphology. Figure \ref{ngc507_centbox} revealed the presence
of secondary emission peaks and cavities in the X-ray emission. The weak
dependence of the HRI spectral response on the temperature near 1 keV 
\citep{clar97}, suggests a correlation between surface brightness and
density fluctuations of the hot gas. The comparison of these X-ray features
with the radio maps of Figure \ref{ngc507_radio} supports the scenario where
the hot gas has been displaced by the pressure of the radio-emitting plasma.
In fact the very good agreement between the shape of the Western radio lobe
and the X-ray cavity indicates that the lobe may have produced a low-density
zone in the gaseous medium, compressing the gas at the edge of the lobe 
\citep[see ][]{clar97}. This interaction is also likely to be responsible for
the N-S elongation of the lobe since the radio plasma expanded avoiding the
denser Western and Southern regions.

The presence of such interactions is not so evident in the Eastern lobe.
This may be due to projection effects and/or to the fact that the Eastern
lobe seems to expand in a less dense environment (the X-ray centroid of the
extended component is displaced by $\sim 20"$=6 kpc South-West of the
nuclear radio source). The Radio/X-ray comparison in left panel of Figure 
\ref{ngc507_radio} shows that a secondary emission peak lies in the center
of the Eastern radio lobe. We followed \cite{Fei95} to calculate the
expected X-ray emission due to Inverse Compton scattering of Cosmic
Background photons into the X-ray band, comparing the minimum energy field
strength $B_{ME}$ with the magnetic field $B_{IC}$ required to produce a
measured ratio of radio to X-ray flux. We used the radio flux for the
Eastern lobe measured by \cite{deRuiter86} and we measured the X-ray flux
density extracting counts from a 25" circle centered on the secondary
emission peak. We found that the expected X-ray emission falls an order of
magnitude below the measured excess for any choice of the involved
parameters, and thus Inverse Compton scattering is unlikely to produce the
emission peak. Conversely, the higher resolution radio contours shown in
right panel of Figure \ref{ngc507_radio} suggest that the bulk of the radio
emission may fall south of the peak, in a low emitting region. In this case
both the secondary emission peak and the SE `tongue' found in the adaptively
smoothed image (Figure \ref{ngc507_csmooth}) may be produced by hot gas
compressed by the radio-emitting plasma, similarly to what seen for the
Western lobe.

The thermal confinement of the radio lobes is supported by the interactions
between the radio and X-ray emitting plasma discussed above. \cite
{Morganti88} studied the pressure balance between the radio lobes and the
surrounding ISM, finding that the thermal pressure greatly exceeded the
radio one ($P_{thermal}/P_{radio}>7$). However their estimate was based on 
\textit{Einstein} data, which resulted in a temperature twice the ROSAT PSPC
value. Taking into account the ROSAT temperature reduces the disagreement by
more than 50\%; the remaining excess may be explained by the departures from
the minimum pressure conditions discussed in their work.

If the secondary peaks are due to hot gas compressed by the radio emitting 
plasma we can estimate the mass and energy contained in such overdensities. 
Assuming that the excess emission is due to spherical bubbles
with radii of 25" (8 kpc) and 17" (5 kpc) for the western and eastern lobe 
respectively we calculate a mean density of $\sim 6-7\times10^{-3}$ cm$^{-3}$
and a total mass of a few $10^8$ M$_\odot$. The lack of energy resolution of 
the HRI data does not allow us to estimate the temperatures of the 
overdensities; however the \textit{Chandra} temperature map obtained by 
\citep{Forman01b} suggests that the bubble temperature is not significantly 
different from the one of the surrounding medium. Assuming $kT\sim 1$ keV we 
obtain a mean pressure $P\sim 2\times 10^{-11}$ erg cm$^{-3}$. Thus the density 
and pressure of the ISM contained in the bubbles are approximately twice those 
of the surrounding gas.
An estimate of the energy required to produce such peaks can be obtained from 
the excess of thermal energy $3nkTV$ contained in the bubbles  with respect to 
the ambient medium. We obtain $\sim 10^{57}$ ergs summing both bubbles. This 
estimate is consistent within a factor of $\sim 2$ with the energy required to 
produce the western cavity $\gamma/(\gamma-1)PV$, which is a good agreement 
considering the uncertainties involved in the calculation.
Since the bubbles are overpressurised they will expand in the surrounding ISM, 
releasing part of their internal energy as mechanical work. Assuming that the 
expansion is adiabatic all the excess internal energy will be transferred to 
the ISM. The energy injected by the central radio source offers a possible 
explanation to the inconsistencies of the cooling flow scenario discussed in 
$\S$ \ref{ngc507_central_discuss}: in fact if radio sources are intermittent 
phenomena with evolutionary timescales of $10^{7}$ yr we obtain an energy 
input in the ISM of $\sim 10^{42}$ erg/s, which is comparable with the energy 
radiated by the X-ray emitting gas within the central 1.5 arcmin of the NGC 507 
halo. This energy input should significantly affect the gas cooling time 
preventing the hot gas to inflow and deposit in the halo core.
This result is in excellent agreement with the simulations performed by 
\cite{Rizza00}. In their models the interaction of the radio lobe with the 
ISM produces X-ray excess emission around the radio lobes of $\sim 20-30\%$ 
such as those observed in NGC 507. Moreover they estimate that the kinetic 
energy of the radio jets is comparable or higher than the energy lost by the 
cooling flow via thermal emission.

Finally we notice that \cite{Canosa99}, using the ROSAT HRI data, found an
upper limit on the X-ray flux of a possible nuclear source that exceeds, by
more than two orders of magnitude, the emission expected from the Radio
data. The extended nature of the central peak implies that the flux of the
nuclear source is much smaller than thought before \citep{Forman01a}, thus
reducing the disagreement between the radio and X-ray data.

\subsection{Discrete Sources}

\label{ngc507_sources_discuss}

\cite{Kim95} revealed the presence of many additional pointlike sources in the NGC 507 halo. The co-added spectrum of these sources was found to be well fitted by
a soft thermal model (kT=0.52) with low abundance (0.0-0.2), leading the
authors to speculate that they could be cooling clumps in the gaseous halo.
As discussed in $\S$ \ref{ngc507_sources} we used the wavelets algorithm to
investigate the nature of discrete sources in the HRI field (Figure \ref
{ngc507_sources_fig}). After excluding NGC 507 (source No.10), NGC 499
(No.11), we cross-correlated the HRI sources (Table \ref{ngc507_source_tab})
with those detected in the PSPC data (Table 2 of \cite{Kim95}). Only 5 out
of the 17 PSPC sources sources falling within the HRI field\footnote{%
Sources No.1,2,19,21 of \cite{Kim95} fall outside or on the edge of the HRI
FOV.} are detected in the HRI data. We compared the fluxes measured by the
two instruments, since any temporal variability would exclude the extended
nature of these sources. As shown in Tables \ref{ngc507_source_tab2} and \ref
{ngc507_source_tab3} two sources varied significantly between 1993 and 1995.
One more object (Source No.7) is likely to be associated with a background
radio source.

In conclusion $\sim 20\%$ of the sources detected by \cite{Kim95} are likely
to be compact objects; this number must be considered as a lower limit since
4 additional HRI sources possess an optical counterpart candidate. This
result is not necessarily in disagreement with the soft spectrum reported by 
\cite{Kim95} since \cite{Sar01} claimed the discovery of supersoft sources
in NGC 4697 (even though their sources are preferentially located near the
galaxy center). The nature of the remaining sources is still uncertain. If
the emission is due to accreting binaries associated with NGC 507 these
systems must host massive black holes since their luminosity would exceed,
by at least an order of magnitude, the Eddinghton luminosity for a 1.4 $%
M_\odot$ Neutron Star. In such case they may be placed in globular clusters,
given the large distances from the galaxy center 
\citep[see for instance][]{Ang01}.
However, we can not exclude the diffuse origin of
many sources since the irregular structure of the gaseous halo and the
filamentary features seen in the halo core suggest that significant density
fluctuations may be present in the hot gas distribution, similar those found
for NGC 1399 \citep{paolillo02}.

\section{Conclusions}

The ROSAT HRI data showed, in agreement with previous studies, a bright
X-ray halo surrounding NGC 507. The large-scale surface brightness
distribution is irregular and more extended in the North direction. The halo
core revealed a complex morphology: there is a main X-ray peak, coincident
with the position of the optical galaxy, and several secondary peaks, due to
the interaction between the radio lobes and the surrounding ISM. 

Modelling the halo surface brightness with a bidimensional double Beta model,
we found that the extended halo centroid is displaced $\sim 22"$ (7 kpc)
South-West of the optical galaxy. This result suggests a different origin
for the central and the extended components. 
The ROSAT and \textit{Chandra} data show that the central X-ray peak
is  produced by dense hot gas in the galaxy core. Even though the cooling time
of the halo within 250" (80 kpc) is shorter than the Hubble time, the mass of
the gas present in the central component falls two orders of magnitude below
the amount expected to be deposited by both homogeneous or inhomogeneous
cooling-flow models. The core luminosity and temperature are instead in good
agreement with the scenario where the central X-ray peak is due to gas of
stellar origin kinetically heated by stellar mass losses. This result is
supported by the solar metallicity found by \cite{Buo00a} and suggests that
the core of bright galaxies is similar to the halo of X-ray compact
early-types studied by \cite{Matsu01}.

The displacement of the cluster halo from the optical galaxy and the
filamentary structures observed in the halo core suggest that the galaxy may
be slowly moving within the group potential. The total mass profile derived
from the bidimensional model shows that a significant amount of dark matter
is required at large radii, if the gas is in hydrostatic equilibrium. The
dark halo extends on group scales and is likely associated with the whole
cluster rather than with the optical galaxy.

Comparing with data in the literature, we find that the separation of
the gaseous halo in two different dynamical ranges (core associated 
to the stellar distribution; external halo tracing the larger potential structures
corresponding to galaxy groups and clusters) is a common feature of many
X-ray extended galaxies. While this result may solve the problem of the
different X-ray properties among X-ray extended and compact galaxies, since
the latters are missing the large-scale component 
\citep{ikebe99,Maki01,Matsu01}, it makes more difficult to explain the
missing deposited mass expected from the short cooling times.

In NGC 507 we found that, if radio sources are intermittent phenomena
with timescales of $10^{7}$ yrs, the amount of energy injected in the ISM is 
large enough to prevent the gas from cooling in the halo core and may explain 
the failure of the cooling flow scenario.

Finally we found that $\sim 20\%$ of the sources detected by \cite{Kim95} in
the NGC 507 halo are due to discrete sources. The nature of the remaining
population is unclear. If associated with NGC 507, they can be either
accreting binaries hosting a massive black hole or density clumps of the
X-ray halo.

~\newline
We acknowledge partial support from the CXC contract NAS8-39073 and NASA
grant NAG5-3584 (ADP), from the MURST PRIN-COFIN 1998-1999 and 2000-2002
(resp. G. Peres) and from the European Social Fund (F.S.E.).\newline
This work is part of the PhD thesis of M. Paolillo at the Palermo University.%
\newline
This research has made use of the NASA/IPAC Extragalactic Database (NED)
which is operated by the Jet Propulsion Laboratory, California Institute of
Technology, under contract with the National Aeronautics and Space
Administration.


\clearpage

\begin{table}[tbp]
\caption{The ROSAT HRI observation of the NGC 507/NGC499 field.}
\label{ngc507observations}
\begin{center}
{\footnotesize 
\begin{tabular}{ll}
&  \\ 
\tableline \tableline name & NGC 507 \\ 
Field center &  \\ 
(R.A., Dec) & 01$^{\mathrm{h}}$23$^{\mathrm{m}}$39$^{\mathrm{s}}$ 33$^\circ$%
15'36'' \\ 
sequence id. & RH600680n00 \\ 
Exp. time & 28310 s \\ 
obs. date & 1995 Jan 14 \\ 
P.I. & D.-W. Kim \\ 
No. OBI & 2 \\ 
Distance\tablenotemark{~(a)} & 65.8 Mpc \\ 
$r_e$\tablenotemark{(b)} & 53 arcsec\\
$N_H$\tablenotemark{(c)} & $5.3\times 10^{20}$ cm$^{-2}$ \\ 
\tableline & 
\end{tabular}
\tablenotetext{(a)}{Estimated assuming $H_0=75$ km s$^{-1}$ Mpc$^{-1}$ and a
velocity of 4934 km s$^{-1}$ \citep{Huchra99}} 
\tablenotetext{(b)}{effective radius \citep{RC3}}
\tablenotetext{(c)}{Galactic
line-of-sight column density \citep{Stark92}} }
\end{center}
\end{table}

\begin{table*}[]
\caption{Best-fit parameters for the bidimensional halo model.}
\label{ngc507_fit_tab}
\begin{center}
{\footnotesize 
\begin{tabular}{lcccccccc}
&  &  &  &  &  &  &  &  \\ 
\tableline \tableline Component & \multicolumn{2}{c}{Center Position} & $r_0$
& $\beta$ & $\epsilon$ & $\theta$ & $\chi^2_\nu$ & $\nu$ \\ 
& R.A. & Dec & (arcsec) &  & $(1-\frac{minor~axis}{major~axis})$ & (rad) & 
& (d.o.f.) \\ 
\tableline Central & 01$^{\mathrm{h}}$23$^{\mathrm{m}}$39.8$^{\mathrm{s}}$ & 
33$^\circ$15'26'' & 0.22$\pm $0.02 & 0.58$\pm $0.01 & 0.0\tablenotemark{(a)}
& 0.0\tablenotemark{(a)} & 0.5 & 73 \\ 
Extended & 01$^{\mathrm{h}}$23$^{\mathrm{m}}$38.7$^{\mathrm{s}}$ & 33$^\circ$%
15'07'' & 49$\pm$ 9 & 0.48$\pm 0.04$ & 0.0\tablenotemark{(a)} & 0.0%
\tablenotemark{(a)} & 0.9 & 776 \\ 
\tableline &  &  &  &  &  &  &  & 
\end{tabular}
\tablenotetext{(a)}{ The parameters $\epsilon$ and $\theta$ have no error
because they were held fixed during the fit (see discussion in text).} 
\tablenotetext{}{{\bf Note} - uncertainties are $1\sigma$ confidence level
for 5 interesting parameters} }
\end{center}
\end{table*}

\begin{table*}[]
\begin{centering}
\caption{Discrete Sources - Results of the Wavelets algorithm \label{ngc507_source_tab}}
\footnotesize
\begin{tabular}{cccrrrccc}
\\
\tableline
\tableline
Source & R.A.         & Dec. & 3$\sigma$ Radius & Counts 	& Max S/N & Count Rate  & $\mathit{f}_{HRI}^{pow.law}$\tablenotemark{(a)} & $\mathit{f}_{HRI}^{RS}$\tablenotemark{(a)} \\
 No.  & \multicolumn{2}{c}{(J2000)} & (arcsec)         &        &         & ($10^{-4}$ Cnts sec$^{-1}$) &    &  \\
\tableline
1  & 1:23:38.6    & +33:21:48	 & 8.7      	& 13$\pm$6  	&  4.3    & 5$\pm$2    	& 35$\pm 14$     				  & 20$\pm 8$ 	   			\\
2  & 1:23:07.5    & +33:15:23	 & 12.2		& 31$\pm$9  	&  7.0    & 11$\pm$3    & 76$\pm 21$     				  & 45$\pm 12$ 	   			\\
3  & 1:23:11.5    & +33:10:58	 & 12.2		& 37$\pm$10  	&  8.0    & 13$\pm$4   	& 90$\pm 28$     				  & 53$\pm 16$ 	   			\\
4  & 1:23:25.0    & +33:25:31	 & 17.3	  	& 47$\pm$13	&  7.6    & 17$\pm$4    & 118$\pm 28$    				  & 69$\pm 16$ 	   			\\
5  & 1:23:43.2    & +33:15:21	 & 17.3		& 33$\pm$11  	&  4.8    & 12$\pm$4   	& 83$\pm 28$     				  & 49$\pm 16$ 	   			\\
6  & 1:22:58.7    & +33:22:43	 & 24.5		& 39$\pm$12  	&  5.2    & 14$\pm$4    & 97$\pm 28$     				  & 57$\pm 16$ 	   			\\
7  & 1:23:44.6    & +33:26:08	 & 24.5		& 33$\pm$11  	&  4.6    & 13$\pm$4    & 90$\pm 28$     				  & 49$\pm 16$ 	   			\\
8  & 1:23:43.4    & +33:32:23	 & 34.6		& 46$\pm$14  	&  4.9    & 21$\pm$6    & 145$\pm 42$    				  & 65$\pm 20$ 	   			\\
9  & 1:23:35.8    & +33:15:03	 & 34.6		& 104$\pm$28  	&  6.1    & 37$\pm$10   & 256$\pm 69$    				  & 151$\pm 41$ 	   			\\
\tableline
\end{tabular}
\tablenotetext{(a)}{ Flux in the 0.1-2.4 KeV band in units of $10^{-15}$ erg s$^{-1}$ cm$^{-2}$.}
\end{centering}
\end{table*}

\begin{table*}[]
\begin{centering}
\caption{Discrete Sources - Comparison with PSPC results and catalogues \label{ngc507_source_tab2}}
\footnotesize

\begin{tabular}{ccccccc}
\\
\tableline
\tableline
Source 	& PSPC\tablenotemark{(a)} & $\mathit{f}_{PSPC}^{RS}$\tablenotemark{(a,b)}& $\mathit{f}_{HRI}^{RS}$\tablenotemark{(b)}& ~~~Time Var.	   & Opt.id. & NED obj.\\
 No.   	&  No.         	       	&     &  											& long-term\tablenotemark{(c)} 		 					     	   &	     & \\
\tableline
1  	& -       		& -     				 & 20$\pm 8$					      & -		   & n       & - \\
2  	& 8       		& $21\pm 10$    			 & 45$\pm 12$					      & n		   & y       & - \\
3  	& 10      		& $171\pm 20$   			 & 53$\pm 16$					      & y		   & y       & - \\
4  	& 12      		& $91\pm 12$    			 & 69$\pm 16$					      & n		   & y       & - \\
5  	& -       		& -     				 & 49$\pm 16$					      & -		   & n       & - \\
6  	& 7       		& $81\pm 12$    			 & 57$\pm 16$					      & n		   & y       & - \\
7  	& 13      		& $43\pm 10$    			 & 49$\pm 16$					      & n		   & y       & NVSS J012345+332554 \\
8  	& -       		& -     				 & 65$\pm 20$					      & -		   & n       & - \\
9  	& -       		& -     				 & 151$\pm 41$  				      & -		   & n       & - \\

\tableline
\end{tabular}

\tablenotetext{(a)}{ As reported in Table 2 of \cite{Kim95}}
\tablenotetext{(b)}{ Flux in the 0.1-2.4 KeV band in units of $10^{-15}$ erg s$^{-1}$ cm$^{-2}$.}
\tablenotetext{(c)}{ Difference between PSPC and HRI fluxes: `y' means variable source ($>3\sigma$), `n' means no variability detected.}
\end{centering}
\end{table*}

\begin{table}[]
\caption{PSPC-HRI Sources Comparison - Upper limits from HRI data }
\label{ngc507_source_tab3}
\begin{center}
{\footnotesize 
\begin{tabular}{ccccc}
&  &  &  &  \\ 
\tableline \tableline PSPC\tablenotemark{(a)} & $\mathit{f}_{PSPC}^{RS}$%
\tablenotemark{(a,b)} & $\mathit{f}_{HRI}^{RS}$ & $\mathit{f}_{HRI}^{\mathrm{%
RS}}$ & Time Var. \\ 
No. &  &  & $3\sigma$ upper~limit & long-term\tablenotemark{(c)} \\ 
\tableline 
3 & $79\pm 15$ & $14\pm 25$ & 90 & - \\ 
4 & $36\pm 10$ & $21\pm 24$ & 93 & - \\ 
5 & $61\pm 14$ & $40\pm 24$ & 111 & - \\ 
6 & $300\pm 27$ & $116\pm 42$ & 243 & y \\ 
9 & $45\pm 11$ & $8\pm 23$ & 77 & - \\ 
11 & $42\pm 11$ & $14\pm 24$ & 86 & - \\ 
14 & $38\pm 10$ & $32\pm 25$ & 108 & - \\ 
15 & $32\pm 10$ & $9\pm 21$ & 72 & - \\ 
16 & $28\pm 10$ & $9\pm 22$ & 75 & - \\ 
17 & $51\pm 11$ & $11\pm 22$ & 78 & - \\ 
18 & $46\pm 13$ & $67\pm 23$ & 135 & - \\ 
20 & $60\pm 14$ & $19\pm 21$ & 83 & - \\ 
\tableline &  &  &  & 
\end{tabular}
\tablenotetext{(a)}{As reported in Table 2 of \cite{Kim95}} 
\tablenotetext{(b)}{ Flux in the 0.1-2.4 KeV band in units of $10^{-15}$ erg
s$^{-1}$ cm$^{-2}$.} 
\tablenotetext{(c)}{\small Difference between PSPC flux
and HRI upper limit: `y' means variable source ($>3\sigma$), `-' means no
variability detected.} }
\end{center}
\end{table}

\begin{table*}[]
\caption{Best-fit spectral parameters for the central component from \textit{%
Chandra} ACIS-S data. }
\label{ngc507_spectral_tab}{\footnotesize 
\centerline{
\begin{tabular}{l|cccc|cccc}
\tableline
\tableline
\bf{Model} & \multicolumn{4}{c|}{($r < 10$ arcsec)} & \multicolumn{4}{c}{($r < 5$ arcsec)}\\
& kT & Abundance & nH & $\chi^2_{red}/\nu$ & kT & Abundance & nH & $\chi^2_{red}/\nu$\\
& (keV) & ($Z/Z_\odot$) & ($10^{20}$ cm$^{-2}$) & & (keV) & ($Z/Z_\odot$) & ($10^{20}$ cm$^{-2}$) & \\
\tableline
Raymond-Smith & $0.82\pm 0.03$ & $0.48^{+0.29}_{-0.16}$ & $16\pm 5		 $ & 0.89/60 & $0.74\pm 0.05$ & $0.48^{-0.21}_{+0.56}$ & $16^{+5}_{-4}  	$ & $0.74/39$\\
Raymond-Smith & $0.84\pm 0.02$ & $1\tablenotemark{(a)}$ & $11\pm 4		 $ & 0.91/61 & $0.77\pm 0.04$ & $1\tablenotemark{(a)}$ & $15^{+8}_{-6}  	$ & $0.75/40$\\
Raymond-Smith & $0.86\pm 0.01$ & $1.3^{+0.8}_{-0.4}   $ & $5.3\tablenotemark{(a)}$ & 0.96/61 & $0.81\pm 2   $ & $1.9^{-0.9}_{+5.4}   $ & $5.3\tablenotemark{(a)}$ & $0.81/40$\\
MekaL         & $0.73\pm 0.04$ & $0.39^{+0.21}_{-0.12}$ & $21\pm 5		 $ & 0.78/60 & $0.68\pm 0.04$ & $0.8^{+1.5}_{-0.4}   $ & $21\pm 6		$ & $0.62/39$\\
MekaL         & $0.77\pm 0.04$ & $1\tablenotemark{(a)}$ & $17\pm 6		 $ & 0.83/61 & $0.68\pm 0.04$ & $1\tablenotemark{(a)}$ & $21^{+7}_{-6}  	$ & $0.60/40$\\
MekaL         & $0.81\pm 0.02$ & $1.53^{+1.2}_{-0.5}  $ & $5.3\tablenotemark{(a)}$ & 0.94/61 & $0.74\pm 0.04$ & $3_{-1.4}^{+123}     $ & $5.3\tablenotemark{(a)}$ & $0.74/40$\\
\tableline
\end{tabular}
\tablenotetext{(a)}{Fixed}
\tablenotetext{}{\small {\bf Note} - Errors are $1\sigma$ for all interesting parameters.}
} 
}
\end{table*}

\clearpage


\newpage

\clearpage

\begin{thebibliography}{} 

\bibitem[Angelini, Loewenstein \& Mushotzky(2001)]{Ang01} Angelini, L., Loewenstein, M., \& Mushotzky, R. F. 2001, ApJ, 557, L35  

\bibitem[Athey et al.(2002)]{Athey02}  Athey, A., Bregman, J., Bregman, J., Temi, P., \& Sauvage, M. 2002, ApJ, 571, 272

\bibitem[Barton, de Carvalho \& Geller(1998)]{Barton98}  Barton, E. J., de
Carvalho, R. R., \& Geller, M. J. 1998, AJ, 116, 1573

\bibitem[Brighenti \& Mathews(1997)]{Brighenti97}  Brighenti, F., \&
Mathews, W. G. 1997, ApJ, 486L, 83

\bibitem[Brown \& Bregman(1998)]{Brown98}  Brown, B. A., \& Bregman, J. N.
1998, ApJ, 495L, 75

\bibitem[Brown \& Bregman(2000)]{Brown00}  Brown, B. A., \& Bregman, J. N.
2000, ApJ, 539, 592

\bibitem[Buote(1999)]{Buo99}  Buote, D. A. 1999, MNRAS, 309, 685

\bibitem[Buote(2000a)]{Buo00a}  Buote, D. A. 2000a, ApJ, 539, 172

\bibitem[Buote(2000b)]{Buo00b}  Buote, D. A. 2000b, ApJ, 544, 242

\bibitem[Caon, Capaccioli \& D'Onofrio(1994)]{Caon94}  Caon, N., Capaccioli,
M., \& D'Onofrio, M. 1994, A\&AS, 106, 199

\bibitem[Canosa et al.(1999)]{Canosa99}  Canosa, C. M., Worrall, D. M.,
Hardcastle, M. J., \& Birkinshaw, M. 1999, MNRAS, 310, 30

\bibitem[Clarke, Harris \& Carilli(1997)]{clar97}  Clarke, D.A., Harris,
D.E., \& Carilli, C.L. 1997, MNRAS, 284, 981

\bibitem[Colla(1975)]{Colla75}  Colla, G., Fanti, C., Fanti, R., Gioia, I.,
Lari, C., Lequeux, J., Lucas, R., \& Ulrich, M. H. 1975, A\&AS, 20, 1

\bibitem[Damiani et al.(1997a)]{Dam97a}  Damiani, F., Maggio, A., Micela,
G., Sciortino, S. 1997a, ApJ, 483, 350

\bibitem[Damiani et al.(1997b)]{Dam97b}  Damiani, F., Maggio, A., Micela,
G., Sciortino, S. 1997b, ApJ, 483, 370

\bibitem[David et al.(1996)]{Dav96}  David, L. P., Harden, Jr., F. R.,
Kearns, K. E., \& Zombeck, M. V. 1996, The ROSAT High Resolution Imager
(HRI), Calibration Report 1996 February, revised, U.S. ROSAT Science Data
Center, SAO, Cambridge, MA

\bibitem[David et al.(1994)]{Dav94}  David, L. P., Jones, C., Forman, W., \&
Daines, S. 1994, ApJ, 428, 544

\bibitem[D'Ercole, Recchi \& Ciotti(2000)]{D'Ercole00}  D'Ercole, A.,
Recchi, S., \& Ciotti, L. 2000, ApJ, 533, 799 (DRC)

\bibitem[de Ruiter et al.(1986)]{deRuiter86}  de Ruiter, H. R., Parma, P.,
Fanti, C., \& Fanti, R. 1986, ApJS, 65, 111

\bibitem[de Vaucouleurs et al.(1991)]{deVau91}  de Vaucouleurs, G., de
Vaucouleurs, A., Corwin, H. G. Jr., Buta, R. J., Paturel, G., \& Fouqu\'{e},
P. 1991, Third Reference Catalogue of Bright Galaxies (New York: Springer)
(RC3)

\bibitem[Fabbiano(1989)]{Fab89}  Fabbiano, G. 1989, ARA\&A, 27, 87

\bibitem[Fabian(1994)]{Fabian94}  Fabian, A. C. 1994, ARA\&A, 32, 277

\bibitem[Fabian(2001)]{Fabian01} Fabian, A. C., Mushotzky, R. F., Nulsen, P. E. J., \& Peterson, J. R. 2001, MNRAS, 321, L20

\bibitem[Fabricant, Lecar \& Gorenstein(1980)]{Fabr80}  Fabricant, D.,
Lecar, M., \& Gorenstein, P. 1980, ApJ, 241, 552

\bibitem[Fanti et al.(1987)]{Fanti87}  Fanti, C., Fanti, R., De Ruiter, H.
R., Parma, P. 1987, A\&AS, 69, 57

\bibitem[Feigelson et al.(1995)]{Fei95}  Feigelson, E. D.,
Laurent-Muehleisen, S. A., Kollgaard, R. I., Fomalont, E. B. 1995, ApJ,
449L, 149

\bibitem[Forman et al.(2001a)]{Forman01a}  Forman, W., Markevitch, M.,
Jones, C., Vikhlinin, A., \& Churazov, E. 2000, astro-ph/0110087

\bibitem[Forman et al.(2001b)]{Forman01b}  Forman, W., Jones, C.,
Markevitch, M., Vikhlinin, A., \& Churazov, E., contribution to the
''Lighthouses of the Universe Conference'', 6-10 August 2001, Garching bei
Munchen, Germany, astro-ph/0111526

\bibitem[Gonz\'{a}lez-Serrano \& Carballo(2000)]{Gonzalez00}  Gonz\'{a}%
lez-Serrano, J. I., \& Carballo, R. 2000, A\&AS, 142, 353

\bibitem[Harris(1999)]{Har99}  Harris, D. E. 1999, in Proceedings of the
colloquium ``The Most Distant Radio Galaxies'', Amsterdam, Royal
Netherlands; eds: H. J. A. R\"{o}ttgering, P. N. Best and Lehnert

\bibitem[Harris et al.(1998)]{Harris98}  Harris, D. E., Silverman, J. D.,
Hasinger, G., \& Lehmann, I. 1998, A\&AS, 133, 431

\bibitem[Hasinger et al.(1993)]{Has93}  Hasinger, G., Burg, R., Giacconi,
R., Hartner, G., Schmidt, M., Trumper, J., Zamorani, G., 1993, A\&A, 275, 1

\bibitem[Huchra, Vogeley \& Geller(1999)]{Huchra99}  Huchra, J. P., Vogeley,
M, S., \& Geller, M. J. 1999, ApJS, 121, 287

\bibitem[Ikebe et al.(1999)]{ikebe99}  Ikebe, Y., Makishima, K., Fukazawa,
Y., Tamura, T., Xu, H., Ohashi, T., \& Matsushita, K. 1999, ApJ, 525, 58

\bibitem[Kaastra(1992)]{Kaastra92}  Kaastra, J.S. 1992, ''An X-Ray Spectral
Code for Optically Thin Plasmas'', Internal SRON-Leiden Report, updated
version 2.0

\bibitem[Kaastra et al.(2001)]{Kaa01}  Kaastra, J. S., Ferrigno, C., Tamura,
T., Paerels, F. B. S., Peterson, J. R., \& Mittaz, J. P. D. 2001, A\&A,
365L, 99

\bibitem[Kim \& Fabbiano(1995)]{Kim95}  Kim, D. W., \& Fabbiano, G. 1995,
ApJ, 441, 182

\bibitem[Kim, Fabbiano \& Trinchieri(1992a)]{Kim92}  Kim, D.-W., Fabbiano,
G., \& Trinchieri, G. 1992a, ApJS, 80, 645

\bibitem[Kim, Fabbiano \& Trinchieri(1992b)]{KFT92}  Kim, D.-W., Fabbiano,
G., \& Trinchieri, G. 1992b, ApJ, 393, 134

\bibitem[Kriss et al.(1983)]{kriss83}  Kriss, G. A., Cioffi, D. F., \&
Canizares, C. R. 1983, ApJ, 272, 439

\bibitem[Liedahl, Osterheld \& Goldstein(1995)]{Liedahl95}  Liedahl, D.A.,
Osterheld, A.L., and Goldstein, W.H. 1995, ApJ, 438L, 115

\bibitem[Mackie et al.(1996)]{mack96}  Mackie, G., et al. 1996, in ASP
Conference Series, Vol. 101, ``Astronomical Data Analysis Software and
Systems V'', G. H. Jacoby and J. Barnes eds.

\bibitem[Makishima et al.(2001)]{Maki01}  Makishima, K., et al. 2001, PASJ,
53, 401

\bibitem[Matsumoto et al.(1997)]{Matsumo97}  Matsumoto, H., Koyama, K.,
Awaki, H., Tsuru, T., Loewenstein, M., \& Matsushita, K. 1997, ApJ, 482, 133

\bibitem[Matsushita(2001)]{Matsu01}  Matsushita, K. 2001, ApJ, 547, 693

\bibitem[Matsushita et al.(2002)]{Matsu02}  Matsushita, K., Belsole, E.,
Finoguenov, A., \& B\"{o}hringer, H. 2002, A\&A, 381,21

\bibitem[Matsushita et al.(1998)]{Matsu98}  Matsushita, K., Makishima, K.,
Ikebe, Y., Rokutanda, E., Yamasaki, N., Ohashi, T. 1998, ApJ, 499L, 13

\bibitem[Matsushita, Ohashi \& Makishima(2000)]{Mat00}  Matsushita, K.,
Ohashi, T., \& Makishima, K. 2000, PASJ, 52, 685

\bibitem[Merrifield(1998)]{Merrifield98}  Merrifield, M. R. 1998, MNRAS,
294, 347

\bibitem[Mewe, Gronenschild \& van den Oord(1985)]{Mewe85}  Mewe, R.,
Gronenschild, E. H. B. M., \& van den Oord, G. H. J. 1985, A\&AS, 62, 197

\bibitem[Mewe, Lemen \& van den Oord(1986)]{Mewe86}  Mewe, R., Lemen, J. R.,
\& van den Oord, G. H. J. 1986, A\&AS, 65, 511

\bibitem[Morganti et al.(1988)]{Morganti88}  Morganti, R, Fanti, R., Gioia,
I. M., Harris, D. E., Parma, P., \& de Ruiter, H. 1988, A\&A, 189, 11

\bibitem[Paolillo et al.(2002)]{paolillo02}  Paolillo, M., Fabbiano, G.,
Peres, G., Kim, D.-W. 2002, ApJ, 565, 883

\bibitem[Parma et al.(1986)]{parma86}  Parma, P., de Ruiter, H. R., Fanti,
C., \& Fanti, R. 1986, ApJS, 64, 135

\bibitem[Peterson et al.(2001)]{Pet01}  Peterson, J. R., et al. 2001, A\&A,
365L, 104

\bibitem[Prugniel \& Simien(1996)]{prugniel96}  Prugniel, P., \& Simien, F.
1996, A\&A, 309, 749

\bibitem[Rangarajan et al.(1995)]{rang95}  Rangarajan, F. V. N., Fabian, A.
C., Forman, W. R., \& Jones, C. 1995, MNRAS, 272, 665 (RFFJ)

\bibitem[Raymond \& Smith(1977)]{RS77}  Raymond, J. C., \& Smith, B. W.
1977, ApJS, 35, 419

\bibitem[Rizza et al.(2000)]{Rizza00} Rizza, E., Loken, C., Bliton, M., Roettiger, K., Burns, O., \& Owen, F.N. 2000, AJ, 119, 21

\bibitem[Sakai, Giovanelli \& Wegner(1994)]{Sakai94}  Sakai, S., Giovanelli,
R., \& Wegner, G. 1994, AJ, 108, 33

\bibitem[Sarazin(1988)]{Sar88}  Sarazin, C. L. 1988, \textit{X-ray Emission
from Clusters of Galaxies}, ed. Cambridge University Press

\bibitem[Sarazin, Irwin \& Bregman(2001)]{Sar01}  Sarazin, C. L., Irwin, J.
A., \& Bregman, J. N. 2001, ApJ, 556, 533

\bibitem[Sarazin \& White(1987)]{Sar87}  Sarazin, C. L., \& White, R. E.
1987, ApJ, 320, 32

\bibitem[Snowden et al.(1994)]{Snow94}  Snowden, S. L., McCammon, D.,
Burrows, D. N., \& Mendenhall, J. A. 1994, ApJ, 424, 714 (SMB)

\bibitem[Stark et al.(1992)]{Stark92}  Stark, A. A., Gammie, C. F., Wilson,
R. W., Bally, J., Linke, R. A., Heiles, C., \& Hurwitz, M. 1992, ApJS, 79, 77

\bibitem[Tamura et al.(2001)]{Tam01}  Tamura, T., et al. 2001, A\&A 365L, 87

\bibitem[Wegner, Haynes \& Giovanelli(1993)]{Wegner93}  Wegner, G., Haynes,
M. P., Giovanelli, R. 1993, AJ, 105, 1251

\bibitem[de Vaucouleurs et al.(1991)]{RC3} de Vaucouleurs, G., de Vaucouleurs, A., Corwin, H. G., Buta, R. J., Paturel, G., Fouque, P., "Third Reference Catalogue of Bright Galaxies (RC3)", Springer-Verlag, New-York (1991)

\end{thebibliography}
\end{document}